\documentclass{aastex63}
\newcommand{\MyFigA}{\ref{MyFigA}}
\newcommand{\MyFigB}{\ref{MyFigB}}
\newcommand{\MyFigC}{\ref{MyFigC}}
\newcommand{\MyFigD}{\ref{MyFigD}}
\newcommand{\MyFigG}{\ref{MyFigG}}
\newcommand{\MyFigE}{\ref{Appendix:MyFigE}}
\newcommand{\MyFigF}{\ref{Appendix:MyFigF}}
\newcommand{\MyTabA}{\ref{MyTabA}}
\newcommand{\MyTabB}{\ref{MyTabB}}
\newcommand{\MyTabC}{\ref{MyTabC}}
\newcommand{\MyTabD}{\ref{Appendix:MyTabD}}
\begin{document}
\title{GRB~200829A: External Shock Origin of the Very Early Prompt Emission?}
\correspondingauthor{Da-Bin Lin, Rui-Jing Lu}
\email{lindabin@gxu.edu.cn, luruijing@gxu.edu.cn}
\author{Jing Li}
\affil{Laboratory for Relativistic Astrophysics, Department of Physics, Guangxi University, Nanning 530004, China}
\author{Da-Bin Lin}
\affil{Laboratory for Relativistic Astrophysics, Department of Physics, Guangxi University, Nanning 530004, China}
\author{Rui-Jing Lu}
\affil{Laboratory for Relativistic Astrophysics, Department of Physics, Guangxi University, Nanning 530004, China}
\author{Lu-Yao Jiang}
\affil{Key Laboratory of Dark Matter and Space Astronomy, Purple Mountain Observatory, Chinese Academy of Sciences, Nanjing 210034, China}
\affil{School of Astronomy and Space Science, University of Science and Technology of China, Hefei, Anhui 230026, China}
\author{Wen-Qiang Liang}
\affil{Laboratory for Relativistic Astrophysics, Department of Physics, Guangxi University, Nanning 530004, China}
\author{Zhi-Lin Chen}
\affil{Laboratory for Relativistic Astrophysics, Department of Physics, Guangxi University, Nanning 530004, China}
\author{Xiao-Yan Li}
\affil{Laboratory for Relativistic Astrophysics, Department of Physics, Guangxi University, Nanning 530004, China}
\author{Xiang-Gao Wang}
\affil{Laboratory for Relativistic Astrophysics, Department of Physics, Guangxi University, Nanning 530004, China}
\author{En-Wei Liang}
\affil{Laboratory for Relativistic Astrophysics, Department of Physics, Guangxi University, Nanning 530004, China}
\begin{abstract}
Long-duration GRB~200829A was detected by \emph{Fermi}-GBM and \emph{Swift}-BAT/XRT,
and then rapidly observed by other ground-based telescopes.
It has a weak $\gamma$-ray emission in the very early phase and followed by a bright spiky $\gamma$-ray emission pulse.
The radiation spectrum of the very early emission is best fitted by a power-law function with index $\sim -1.7$.
However, the bright spiky $\gamma$-ray pulse,
especially the time around the peak,
exhibits a distinct two-component radiation spectra,
i.e., Band function combined with a blackbody radiation spectrum.
We infer the photospheric properties and reveal a medium magnetization at photospheric position
by adopting the initial size of the outflow as $r_0=10^9$~cm.
It implies that Band component in this pulse may be formed during the dissipation of magnetic field.
The power-law radiation spectra found in the very early prompt emission
may imply the external-shock origination of this phase.
Then, we perform Markov Chain Monte Carlo method fitting on the light-curves of this burst,
where the jet corresponding to the $\gamma$-ray pulses at around $20$~s
is used to refresh the external-shock.
It is shown that the light-curves of very early phase and X-ray afterglow after $40$~s,
involving the X-ray bump at around $100$~s,
can be well modelled in the external-shock scenario.
For the obtained initial outflow,
we estimate the minimum magnetization factor of the jet
based on the fact that the photospheric emission of this jet is missed in the very early phase.
\end{abstract}

\keywords {Gamma-ray bursts (629)}
%%%%%%%%%%%%%%%%%%%%%%%%%%%%%%%%%%%%%%%%%%%%%%%%%%%%%%%%%%%%%%%%%%%%%%%%%%%%%%%%%%%
%%%%%%%%%%%%%%%%%%%%%%%%%%%%%%%%%%%%%%%%%%%%%%%%%%%%%%%%%%%%%%%%%%%%%%%%%%%%%%%%%%%
\section{Introduction}\label{Sec:Intro}
Theoretically, it is generally believed that gamma-ray bursts (GRBs)
originated from collapse of massive stars or mergers of double compact stars (e.g., \citealp{Colgate_SA-1974-ApJ.187.333C, Paczynski_B-1986-ApJ.308L.43P, Eichler_D-1989-Livio_M-Natur.340.126E, Narayan_R-1992-Paczynski_B-ApJ.395L.83N, Woosley_SE-1993-ApJ.405.273W, MacFadyen_AI-1999-Woosley_SE-ApJ.524.262M, Piran_T-2004-RvMP.76.1143P, Zhang_B-2004-Meszaros_P-IJMPA.19.2385Z, Woosley_SE-2006-Bloom_JS-ARA&A.44.507W, Kumar_P-2015-Zhang_B-PhR.561.1K}).
Observationally, GRBs generally appear as a brief and intense $\gamma$-rays followed by a long-lived afterglow emission.
The prompt $\gamma$-rays are highly variable with a duration from millisecond to thousands of seconds.
The observational spectra are usually well fitted by an
empirical function, characterized by a smoothly joint broken power-law
function, the so-called Band function (\citealp{Band_D-1993-Matteson_J-ApJ.413.281B})
or a quasi-thermal spectral component
appear in the spectra of some GRBs.
The previous observations demonstrated that
thermal components exhibit different observational properties.
They either can be detected during the entire duration of the
prompt emission (e.g., \citealp{Ghirlanda_G-2013-Pescalli_A-MNRAS.432.3237G}) or
may be only found at the beginning of the burst duration, and
subsequently appear with a nonthermal component.
The detection of a diversified spectral characteristic shows
that GRB ejecta may have a diverse jet composition.
It may be neither fully matter-dominated ejecta nor fully magnetized
outflows. More realistically, GRB outflows are likely to be a
hybrid jet, which carries the two components simultaneously and
launches at the central engine (e.g., \citealp{Gao_H-2015-Zhang_B-ApJ.801.103G}).
The light-curves of afterglow emission usually can be
decomposed into four power-law segments,
i.e., an initial steep decay, a shallow decay, a normal decay, and a late steeper decay,
sometimes accompanied by one or several flares (\citealp{Zhang_B-2006-Fan_YZ-ApJ.642.354Z, Nousek_JA-2006-Kouveliotou_C-ApJ.642.389N}).
It is commonly believed that
the multi-wavelength afterglow is mainly from the external shock,
which is formed during a relativistic jet propagating in the circum-burst medium (e.g., \citealp{Meszaros_P-1997-Rees_MJ-ApJ.476.232M}).
However, the origin of the prompt $\gamma$-rays is not well understood.
The prompt $\gamma$-rays may be from the internal shock in an erratic relativistic fireball,
a dissipative photosphere, a Poynting-flux dominated jet,
or even an external shock (e.g., \citealp{Rees_MJ-1992-Meszaros_P-MNRAS.258P.41R, Meszaros_P-1993-Rees_MJ-ApJ.405.278M, Rees_MJ-1994-Meszaros_P-ApJ.430L.93R, Giannios_D-2008-A&A.480.305G, Beloborodov_AM-2010-MNRAS.407.1033B, Vurm_I-2011-Beloborodov_AM-ApJ.738.77V, Zhang_B-2011-Yan_HR-ApJ.726.90Z, Burgess_JM-2016-Begue_D-ApJ.822.63B, Huang_LY-2018-Wang_XG-ApJ.859.163H}).

It is not a new idea that the prompt $\gamma$-rays of GRBs originate from the external shock.
\cite{Burgess_JM-2016-Begue_D-ApJ.822.63B} have shown that the prompt emission of
GRB~141028A is very likely originated from an external shock.
\cite{Huang_LY-2018-Wang_XG-ApJ.859.163H} suggested that GRB~120729A is an external shock origin
for both the prompt $\gamma$-ray emission and afterglow.
They also systematically investigate single pulse GRBs
in the \emph{Swift}'s GRBs, and find that a small fraction of GRBs (GRBs 120729A, 051111, and 070318)
are likely to originate from an external shock for both the prompt $\gamma$-ray emission and afterglow.
However, \cite{Huang_LY-2018-Wang_XG-ApJ.859.163H} focuses on the bursts appearing as a single pulse
from the prompt emission to its afterglow.
In fact, the central engine of GRBs may re-activity and launch relativistic ejecta several times.
The late launched ejecta may be observed as flares in the afterglow and
interact with the external shock at a later period.
The burst GRB~200829A maybe in the above scenarios.
GRB~200829A was detected by \emph{Fermi}-GBM and \emph{Swift}-BAT/XRT,
and the light-curve of prompt emission is composed of an initial very early weak emission
(with a duration $\sim 5$~s) followed by a bright spiky $\gamma$-ray pulse with a duration
$\sim 10$~s. We find that the spectra in the
$\gamma$-ray pulse of GRB~200829A exhibits a distinct two-component,
i.e., Band function combined with a blackbody radiation spectrum,
especially in the peak time.
It means that the thermal component should be indeed existence, and GRB~200829A outflows are likely to be a
hybrid jet. What's more, the radiation spectrum in its very early phase can be fitted with power-law spectral model with index $\sim -1.7$,
which may be an indication of the origin of an external-forward shock.
The central engine of GRB~200829A may re-activity and launch jets at different times,
resulting in the bright spiky $\gamma$-ray pulses when jets collide with each other.

In this paper, we present a detailed analysis of $\gamma$-rays and
X-ray emission from the long GRB~200829A detected by
\emph{Fermi} and \emph{Swift}.
The paper is organized as follows.
In Section~\ref{Sec:Observation}, we introduce the observations and light-curves features of GRB~200829A.
In Section~\ref{Sec:Analysis}, the detailed analysis and results of GRB~200829A are performed.
In this section, we also analyzed the other properties of GRB~200829A in different phase.
In Section~\ref{Sec:Conclusions}, the summary and discussions are presented.

\section{Observations and data reduction}\label{Sec:Observation}
The long GRB~200829A was first detected by Fermi Gamma-Ray Burst Monitor (GBM) at $13:58:14.66$ UT ($T_0$) on 2020 August
29 with duration $T_{90}\sim 6.9~\rm s$ (\citealp{Lesage_S-2020-Meegan_C-GCN.28326.1L}).
In addition to the \emph{Fermi}-GBM, \emph{Swift}-BAT triggered the burst
at $13:59:34$ UT on 2020 August 29 (\citealp{Palmer_DM-2020-Barthelmy_SD-GCN.28325.1P})
and \emph{Swift}-XRT began to observe the burst at 128.7~s after
the BAT trigger (\citealp{Gropp_JD-2020-Kennea_JA-GCN.28317.1G}).
\cite{Oates_SR-2020-Kuin_NPM-GCN.28338.1O} created a SED at 900~s after the BAT trigger
and found a photometric redshift of $z=1.25 \pm 0.02$ for this burst.
The optical afterglow is detected on first two days after the GRB trigger (\citealp{Pozanenko_A-2020-Pankov_N-GCN.28359.1P}).
In the left panels of Figure~{\MyFigA}, we show the light-curves of prompt $\gamma$-rays
and afterglows of GRB~200829A with respect to the Fermi trigger.
The inset in the upper part of this panel
shows the light-curves of prompt emission
based on the Fermi observation in the linear spaces.
Here, the Fermi data are from the Fermi Science Support Center\footnote{https://fermi.gsfc.nasa.gov/ssc/data/access/}
and a GBM light-curve and source spectra are extracted from the TTE (Time-Tagged-Events) data by using a python source package named $gtBurst$\footnote{https://github.com/giacomov/gtburst},
the BAT/XRT data are taken from the UK Swift Science Data Center\footnote{http://www.swift.ac.uk/burst\_analyser/00993768/},
and the optical data of GRB~200829A are from \cite{Siegel_MH-2020-Gropp_JD-GCN.28307.1S,Pozanenko_A-2020-Reva_I-GCN.28308.1P, Lipunov_V-2020-Kornilov_V-GCN.28309.1L, Kuin_NPM-2020-Siegel_MH-GCN.28311.1K, Lipunov_V-2020-Kornilov_V-GCN.28315.1L, Hentunen_VP-2020-Nissinen_M-GCN.28318.1H, Moskvitin_AS-2020-Aitov_VN-GCN.28322.1M, Zhu_ZP-2020-YFu_S-GCN.28324.1Z, Moskvitin_AS-2020-Aitov_VN-GCN.28328.1M, Pankov_N-2020-Novichonok_A-GCN.28329.1P, Zhu_ZP-2020-YFu_S-GCN.28330.1Z, Izzo_L-2020-GCN.28331.1I, Volnova_A-2020-Naroenkov_S-GCN.28333.1V, DePasquale_M-2020-GCN.28404.1D, Pozanenko_A-2020-Pankov_N-GCN.28359.1P}.

Based on the light-curves in the left panels of Figure~{\MyFigA}, one can find that the prompt $\gamma$-rays is
dominated by a bright spiky $\gamma$-ray pulses in the period of $t_{\rm obs}\sim[15,30]$~s based on GBM observation,
which is preceded by a small $\gamma$-ray pulse in the period of $t_{\rm obs} \sim[6,10]$~s based on BAT observation.
However, it should be noted that the small $\gamma$-ray pulse in the period of $t_{\rm obs} \sim[6,10]$~s
is not significantly in the light-curve of GBM observation.
Except these two $\gamma$-ray episodes,
there is a significant $\gamma$-ray emission in the very early phase of the prompt emission ($t_{\rm obs}<6$~s)
based on BAT observation.
This can also be found in the right panels of Figure~{\MyFigA},
which shows the GBM light-curve of GRB 200829A without background subtracted (upper panel)
and the signal significance (bottom panel).
One can find that the signal significance in the period of $\sim [0, 10]$~s is higher than $\sim 4 \sigma$, which reveal a significant $\gamma$-ray photons in this period.
In the following section, we present the detailed studies on the spectra and the corresponding physical implications
for the very early phase and the bright spiky $\gamma$-ray pulses.

\section{Detailed Analysis of GRB~200829A and Results}\label{Sec:Analysis}
\subsection{Very early prompt gamma-ray emission }\label{Sec:Early}
For the very early phase of the prompt emission,
the spectral fitting with Band function\footnote{Band function is described as $N(E)=N_0 (E/100{\rm keV})^{\alpha} \exp(-E/E_0)$ for $E \le (\alpha  - \beta )E_0$ and $N(E)=N_0 [(\alpha-\beta)E_0/100{\rm keV}]^{\alpha-\beta}\exp(\beta-\alpha)(E/100{\rm keV})^\beta$ for $E\ge(\alpha-\beta)E_0$, where $N_0$ is the normalization, and $\alpha$, $\beta$, and $E_0$ are parameters in the spectral fittings. The peak photon energy of $E^2N(E)$ is ${E_{\rm{p}}} = (\alpha  + 2){E_0}$.}
reports the values of $\alpha=-1.75\pm 0.09$, $E_0=9976.67\pm51113.36$,
and $\beta=-2.42\pm5.08$ (see the third line of Table~{\MyTabA}).
The values of $E_{\rm 0}$ and $\beta$ could not be well constrained from the spectral fitting.
Then, we perform the spectral analysis of the very early phase
with the power-law (PL) function\footnote{The PL function is described as $N(E)=N_0 (E/1{\rm keV})^{\hat{\Gamma}}$ with ${\hat{\Gamma}}$ being the photon spectral index.}
or cutoff power-law (CPL) function.
Here, the spectral fitting with PL function reports the power-law index $\hat{\Gamma}=-1.79\pm0.06$ (see the second line of Table~{\MyTabA}),
and the spectral fitting with CPL function could not present a well fitting
and the corresponding result is not reported.
The spectral fitting results for the very early prompt emission
with Band function (left panel) and PL function (middle panel) are also shown in Figure~{\MyFigB}.
We note that the values of $\alpha=-1.75\pm 0.09$ and
$\hat{\Gamma}=-1.75\pm0.06$ from the spectral fittings are almost the same.
Here, the value of $\alpha$ can be well constrained in the spectral fitting with Band function.
This fact may imply that
the intrinsic radiation spectrum in this period
may be consistent with a PL spectral model with $\hat{\Gamma} \sim -1.7$ or
a Band function with a break at $\sim10$~MeV and power-law index $\sim -1.7$ in its low-energy regime ($E\lesssim 10$~MeV)\footnote{Please see Appendix~\ref{AppendixA} for a comprehensive analysis about the radiation spectrum in this period.}.

The reasons are as follows.
Firstly, the spectral fitting on such kind of intrinsic radiation spectrum
with a Band function would not provide a well constraint on the value of $E_0$ and thus $\beta$.
This is consistent with our spectral fitting result for this period based on a Band function.
In addition,
a Band function with a break at $\sim10$~MeV,
the power-law index $\sim -1.7$ in its low-energy regime ($E\lesssim 10$~MeV),
and the power-law index $\gtrsim-2.5$ in its high-energy regime ($E\gtrsim 10$~MeV)
can be modelled with a PL function and $\hat{\Gamma}=-1.79$ in \emph{Fermi}-GBM energy band (8~keV-40~MeV).
This is also consistent with our spectral fitting result for this period based on a PL function.
Secondly and importantly, we perform the spectral fitting on the Swift-BAT observation
for this period based on a PL model
and the value of $\hat{\Gamma}=-1.71$ is reported.

We note that such kind of intrinsic radiation spectrum is very different from
the general Band radiation component of GRBs' prompt emission,
of which the value of $\alpha$ is around $-1$ and the break energy $E_0$ is around $400$~keV.
The right panel of Figure~{\MyFigB} shows the relation of $E_{\rm p}$ and $\alpha$
based on the spectral fitting results with a Band function,
where the blue symbols are from the figure~8 of \cite{Poolakkil_S-2021-Preece_R-ApJ.913.60P}
and represent the GOOD sample for time-integrated spectral fits with Band function.
In this panel, the spectral analysis result for the very early phase of the prompt emission
based on Band function is also tentatively shown with pink ``$\bigstar$''
even though the value of $E_0$ could not be well constrained,
and the spectral fitting results of the small $\gamma$-ray pulse ($[5,10]$~s)
or the bright spiky $\gamma$-ray pulses ($[16,26]$~s) with Band function are also showed.
One can find that such kind of radiation spectrum is very different from
the general Band radiation component of GRBs' prompt emission,
involved that of the bright spiky $\gamma$-ray pulses or the small $\gamma$-ray pulse.
Then, we would like to believe that the very early phase of the prompt emission
in this burst may be originated from the other channel
rather than that for the bright spiky $\gamma$-ray pulses or the small $\gamma$-ray pulse.

\subsection{Bright spiky $\gamma$-ray pulse and deriving physical parameters}\label{Sec:Pulse}
There is a bright spiky $\gamma$-ray pulse appearing at $t_{\rm obs}\sim[16,26]$~s after the Fermi trigger.
In order to perform detailed analysis of this pulse,
we divide this pulse into several time intervals with 1~s time span
and perform the spectral fitting on these time intervals with Band function.
The spectral fitting results are reported in Table~{\MyTabB}
and shown in the left panels of Figure~{\MyFigC}.
A distinct multi-component of radiation spectrum is found in several time intervals of this pulse,
e.g., [18, 19]~s.
Then, we also perform the spectral analysis together with Band function and a blackbody radiation component (BB) \footnote{$N_{\textrm{BB}}(E)=\frac{ 8.0525\times K E^2}{(kT)^4 (e^{(E/kT)}-1)}$,
where \emph{kT} is the blackbody temperature keV;
\emph{K} is the \emph{L$_{39}$}/ \emph{D$_{10}^2$}, where \emph{L$_{39}$} is the
source luminosity in units of {10$^{39}$} erg/s and
\emph{D$_{10}$} is the distance to the source in units of 10 kpc.},
i.e., ``Band+BB''.
The spectral fitting results based on Band+BB model are also reported in Table~{\MyTabB}
and shown in the right panels of Figure~{\MyFigC}.
We also estimate the Bayesian Information Criterion (BIC; \citealp{Schwarz_G-1978-AnSta.6.461S})
for the spectral fitting with Band function and that with Band+BB model.
The values of BIC from the spectral analysis are also reported in Table~{\MyTabB}.
The BIC is adopted to evaluate the goodness of the
model fitting, taking into account the model complexity and the different numbers of free parameters.
Generally, the model with a lowest BIC is preferred.
By comparing the values of BIC from the spectral analysis,
one can find that the Band+BB model is preferred
for the radiation spectrum of the time intervals around the peak of the bright spiky $\gamma$-ray pulse.
Since the value of $\Delta$BIC=$\rm BIC_{\rm Band}-BIC_{\rm Band+BB}$ is in the range of 12-25,
it is strong to support a blackbody component in these time intervals\footnote{In the spirit of \cite{KennethP_B-DavidR_A},
the value of $\Delta$BIC can be used as the strength of the evidence
to allow a quick comparison and ranking of candidate hypotheses or models.
For $\Delta \rm BIC=\rm BIC_{\rm A}-\rm BIC_{\rm B}$ with $\rm BIC_{\rm A}>\rm BIC_{\rm B}$,
the strength of the evidence can be summarized as follows:
the situation with $\Delta \rm BIC\leqslant 2$ provides no evidence against the model-A;
the situation with $4\leqslant\Delta \rm BIC\leqslant7$ provides
positive evidence against the model-A;
the situation with $\Delta \rm BIC \geqslant 10$ provides very strong evidence
against the model-A (\citealp{KennethP_B-DavidR_A}).}.

The temperature and flux of the blackbody component,
together with the radius of the jet base (size of the central engine) $r_0$ and $z$,
can provide useful information about the physics of the photosphere.
Meanwhile, due to the presence of Band energy spectrum component,
the jet compositions of GRB~200829A maybe hybrid.
Therefore, following \cite{Gao_H-2015-Zhang_B-ApJ.801.103G},
we estimate the radius and Lorentz factor
of the photosphere based on the blackbody component
found in the period of [18,22]~s by assuming the hybrid outflow of GRB~200829A.
In the calculations, we assume that there is no dissipation below the photosphere and
the radiation efficiency $\sim 52.1\%$ (please see Section~\ref{Sec:Conclusions}).
The results are shown in the left panels of Figure~{\MyFigD},
the blue and olive symbols are the physical quantities calculated based on $r_0=10^8$~cm and $r_0=10^9$~cm,
solid and hollow ``$\bigstar$'' represent the physical parameters $r_{\rm ph}$, $\Gamma_{\rm ph}$, respectively.
It indicates that the values of $r_{\rm ph}$ increases with time
and ${\Gamma}_{\rm ph}$ remains constant for low value of $r_0$
and when $r_0$ is large, it increases and eventually declines.
We also infer the dimensionless entropy $\eta$ and the magnetization factor $\sigma$,
where $\sigma_0$ and $\sigma_{\rm ph}$ are the magnetization factor of the outflow at $r_0$ and $r_{\rm ph}$,
respectively.
The results are shown in the middle panels of Figure~{\MyFigD},
the blue and olive symbols are the same as those in the left panels of Figure~{\MyFigD},
and solid and hollow ``$\bigstar$'' represent the physical parameters $\eta$, $1+\sigma_{\rm ph}$, and $1+\sigma_0$, respectively.
It is shown that the dimensionless entropy $\eta$ fluctuates in the range of 100 to 300.
In addition, the values of $1+\sigma_{\rm ph}$ can be around $5$ if $r_0=10^9$~cm is adopted
and around $1$ if $r_0=10^8$~cm is adopted.
Together with the Band and BB components found in this burst,
the initial radius of the outflow producing the bright spiky $\gamma$-rays
should be around or larger than $10^9$~cm, i.e., $r_0\gtrsim10^9$~cm.
This result is consistent with that found in GRBs with identified photospheric emission,
e.g., GRB~120323A, GRB~131014A and GRB~220426A (e.g., \citealp{Guiriec_S-2013-Daigne_F-ApJ.770.32G,Guiriec_S-2015-Mochkovitch_R-ApJ.814.10G,Deng_LT-2022-Lin_DB-ApJ.934L.22D}).
The non-thermal component in the bright spiky $\gamma$-rays,
i.e., Band component, seem to be formed during the dissipation of the magnetic energy.

\subsection{Afterglow analysis and a self-consistent Paradigm for bursting}\label{Sec:Afterglow}
Following the prompt $\gamma$-ray emission in this burst,
a late bump appears at $t_{\rm obs}>40$~s with a rising in the period of $t_{\rm obs}\sim[40,100]$~s
and a decaying after $t_{\rm obs}\sim100$~s.
It is reasonable to believe that the decaying phase of the late bump
is the normal decay of the external-forward shock.
For the X-ray emission in this phase,
the closure relation (\citealp{Zhang_B-2004-Meszaros_P-IJMPA.19.2385Z})
of $\alpha\approx 3\beta/2$ with $F\propto \nu^{-\beta}t^{ -\alpha}$ can be found,
where the value of $\alpha=1.30 \pm 0.03$ and $\beta=0.80\pm0.05$ are obtained based on the observations of \emph{Swift}.
It reveals that the X-ray emission in this phase is in the spectral regime of $\nu_{\rm m}<\nu<\nu_{\rm c}$
for an external-forward shock in the interstellar medium.

The very early phase of the prompt emission may be originated from the external shock.
The reasons are as follows.
Firstly, we have performed a joint spectral analysis by
combining the observations of \emph{Swift}-BAT and \emph{Fermi}-GBM
for the very early phase of the prompt emission in Section~\ref{Sec:Early}.
The spectral analysis reveals that the very early phase of the prompt emission in this burst may be originated from the other channel rather than that for the small $\gamma$-ray pulse or the bright spiky $\gamma$-ray pulses.
Secondly, the radiation spectrum in this phase is strongly reminiscent of the GRB~120729A,
of which the radiation spectrum in the prompt emission for \emph{Fermi}-GBM energy band
can be well modelled with a PL function and
photon spectral index $\hat{\Gamma}\sim -1.47$\footnote{By performing joint spectral fitting of the \emph{Swift}-BAT and \emph{Fermi}-GBM observations for GRB~120729A, we obtain $\hat{\Gamma}\sim -1.47$ and $\hat{\Gamma}\sim -1.49$ for the period of [0, 10]~s and [1, 2]~s after the Fermi trigger, respectively.}(\citealp{Huang_LY-2018-Wang_XG-ApJ.859.163H}).
Since the light-curve of the prompt emission in GRB~120729A appears as a single long and smooth pulse,
which extends continuously to the X-rays,
it is suggested that both the prompt emission and the afterglows are originated from an external-forward shock (\citealp{Huang_LY-2018-Wang_XG-ApJ.859.163H}).
Thirdly, the spectral index of the very early prompt emission based on \emph{Swift}-BAT and \emph{Fermi}-GBM
observations is almost the same as that of the decaying phase in the late bump based on the \emph{Swift}-XRT observation (see Table~{\MyTabA} and Table~{\MyTabC}).
This is different from that in GRB~120729A,
of which the spectral index in the X-ray energy band evolves from -1.47 in the early phase of the prompt emission to -1.83 in the late phase of afterglow.
It may reveal that the X-rays may in the same spectral regime in GRB~200829A
but in different spectral regime in GRB~120729A for the very early prompt emission and the late phase of afterglow.
Then, we would like to believe that the early phase of prompt emission ($t_{\rm obs}<6$~s) has a same origination
as that of the decaying phase of the late bump,
i.e., they all stem from the external-forward shock.
In addition, the two $\gamma$-ray pulses in the period of $\sim[6,26]$~s should
reflect the re-activity of the central engine of GRB~200829A.

Then, we suggest that the central engine of GRB~200829A may be intermittent
and launch several episode of ejecta separated by a long quiescent interval
(\citealp{Lin_DB-2018-Huang_BQ-ApJ.852.136L}).
The very early phase of the prompt emission originates from the external shock,
which is formed during the propagation of the first launched ejecta in the circum-burst medium.
The later launched ejecta, of which the internal dissipation is responsible for the two $\gamma$-ray pulses,
collide with the formed external shock in the period of $t_{\rm obs}\sim[60,100]$~s.
Then, the energy injection into the external shock is presented in this period
and correspondingly a rising phase appears in the period of $t_{\rm obs}\sim[60,100]$~s.
Based on the above paradigm, we fit the very early prompt emission and the late bump
with an external-forward shock in the ISM (see Appendix~\ref{AppendixB} for detail modeling),
of which the free parameters are the isotropic kinetic energy $E_{\rm k,0}$,
the initial Lorentz factor $\Gamma_{\rm 0}$, the
fraction of shock energy to electron energy $\epsilon_{e}$, the fraction of
shock energy to magnetic field energy $\epsilon_{_B}$,
the interstellar medium density $n_0$, the jet opening angle
$\theta_j$, and $\delta$.
Here, the energy injection rate of the external-forward shock in the period of $[t_{\rm s}, t_{\rm e}]=[20, 100]$~s is described as ${d{E_{{\rm{inj}}}}/d{t_{{\rm{obs}}}}}=E_{\rm k,0}\delta/(t_{\rm e}-t_{\rm s})$
with $\delta$ being a free parameter in out fitting.
In our fitting, a Markov Chain Monte Carlo method based on the emcee Python package (\citealp{Foreman-Mackey_D-2013-Hogg_DW-PASP.125.306F}) is adopted
to search for the best-fit parameter set.
The optimal result is shown in the left panel of Figure~{\MyFigA}
with wine line for X-ray data and blue line for optical data,
and the obtained parameters at the 1$\sigma$ confidence level are $\log_{10} E_{\rm k,0} = 53.65_{-0.07}^{+0.07}$~erg,
$\log_{10}\Gamma_0 = 3.17_{-0.01}^{+0.05}$,
$\log_{10}\epsilon_e = -0.31_{-0.01}^{+0.01}$, $\log_{10}\epsilon_B = -5.15_{-0.19}^{+0.17}$,
$\log_{10} n_0 = 1.27_{-0.18}^{+0.19} \rm\ cm^{-3}$, $p = 2.001_{-0.001}^{+0.002}$,
$\theta_j = 0.09_{-0.01}^{+0.01}$, $\log_{10}\delta = 0.81_{-0.03}^{+0.04}$.
The corresponding posterior probability density functions for the physical parameters are presented in Figure~{\MyFigG}.
From the left panel of Figure~{\MyFigA}, one can find that the external-forward shock with a refreshed phase can well describe both the very early prompt emission and the late bump in the afterglows for GRB~200829A.

\section{Summary and Discussiones}\label{Sec:Conclusions}
Observationally, GRB~200829A appears with a weak $\gamma$-ray emission in the very early phase,
followed by a small $\gamma$-ray pulse at around $6$~s
and a bright spiky $\gamma$-ray pulse at around $20$~s
after the Fermi trigger.
After the bright spiky $\gamma$-ray pulse,
a smooth bump in the X-ray bands appears.
We perform detail spectral analysis on
the very early prompt emission
and the bright spiky $\gamma$-ray pulse.
It reveals that
the very early prompt emission can be well fitted by a power-law spectral model with index $\sim -1.7$.
However, the bright spiky $\gamma$-ray pulse,
especially the time around the pulse peak,
exhibits a distinct two-component, i.e., Band function combined with a blackbody radiation spectrum.
This indicate that the origination of the very early prompt emission and the bright spiky $\gamma$-ray pulse
may be different.
The power-law spectral index of the very early prompt emission is almost
the same as that of the normal decay phase in the X-ray smooth bump,
which is suggested to be originated from the external-forward shock.
Then, we suggest that the central engine of GRB~200829A may be intermittent
and launch several episode of ejecta separated by a long quiescent interval.
The very early phase of the prompt emission originates from the external shock,
which is formed during the propagation of the first launched ejecta in the circum-burst medium.
The later launched ejecta, of which the internal dissipation is responsible for the two $\gamma$-ray pulses,
collide with the formed external shock in the period of $t_{\rm obs}\sim[60,100]$~s.
Then, the energy injection into the external shock is presented in this period
and correspondingly a rising phase appears in the period of $t_{\rm obs}\sim[60,100]$~s.
Based on the above paradigm, we fit the very early prompt emission and the late bump
with an external-forward shock in the ISM based on Markov Chain Monte Carlo method.
It is shown that the light-curves of the very early prompt emission, X-ray afterglow after $40$~s
involving the X-ray bump at around $100$~s,
and the later optical afterglow
can be well modelled in the above paradigm.

We also perform detail study on the jet producing the bright spiky $\gamma$-ray pulse.
Based on the blackbody radiation component found in this pulse,
the magnetization of the jet at the photosphere
is estimated to be $\sim 4$ if the initial size of the fireball $r_0=10^9$~cm is adopted.
Then, the non-thermal component in the bright spiky $\gamma$-rays,
i.e., Band component, seems to be formed during the dissipation of the magnetic energy.
This may lead to a high radiation efficiency of the jet.
With the energy injection in the period of $[20,100]$~s,
the radiation efficiency of the bright spiky $\gamma$-ray pulse
is estimated as $\eta_{\rm \gamma} =
E_{\gamma}/(E_{\gamma} + E_{\rm inj})\sim 52.1\%$,
where ${E_{{\rm{inj}}}}={d{E_{{\rm{inj}}}}/d{t_{{\rm{obs}}}}}\times (t_{\rm e}-t_{\rm s})$
and $E_{\gamma}\approx 1.41\times 10^{54}$~erg is the isotropic energy of the bright spiky $\gamma$-ray pulse.
The obtained high value of radiation efficiency is consistent with the scenario that
the non-thermal component in this pulse is formed during the dissipation of the magnetic energy in the jet.
Besides, the Lorentz factor of the jet at the photosphere is estimated to be around 500
(400) if $r_0=10^8$~cm ($r_0=10^9$~cm) is adopted.
The Lorentz factor of the jet can also be estimated as follows.
The distance of the jet dissipation location $r_{\rm dis}$ relative to the central engine of the burst and the Lorentz factor $\Gamma_{\rm dis}$ of the dissipation region
may be related to the pulse duration $\Delta t_{\rm pulse}$ as
$\Delta t_{\rm pulse}= R_{\rm dis} /(2\Gamma_{\rm dis}^2c)\sim 4$~s (full-width at half maximum).
In addition, the dissipation location should be less than the location of the external shock
at the same observer time, i.e., $R_{\rm dis}\lesssim R_{\rm es,20~s}\sim 4\times10^{16}$~cm,
where $R_{\rm es,20~\rm s}$ is the location of the external shock at the observer time 20~s
and obtained based on the initial fireball
(without energy injection) and Equations~(\ref{Eq:Gamma})-(\ref{Eq:_beta}).
Then, one can have $\Gamma_{\rm dis}\lesssim 408$.
Interesting, the Lorentz factor of the jet producing the bright spiky $\gamma$-ray pulse
can be estimated based on the blackbody radiation component.
We find that the Lorentz factor of the jet is consistent with that
estimated based on the blackbody radiation component in
the bright spiky $\gamma$-ray pulse.
Please see the left panel of Figure~{\MyFigD},
where $\Gamma_{\rm ph}\sim 400$ is obtained if $r_0=10^9$~cm is adopted.

The magnetization of the outflow would affect its photospheric emission
(e.g., \citealp{Zhang_B-2009-Peer_A-ApJ.700L.65Z,Gao_H-2015-Zhang_B-ApJ.801.103G}).
Since the emission of the initial fireball, involving the photospheric emission,
missed in the observation,
the magnetization of the initial fireball would be high.
In the spirit of \cite{Zhang_B-2009-Peer_A-ApJ.700L.65Z},
the outflow with magnetization $\sigma\gtrsim 125$
($\sigma\gtrsim 162$) is required if $r_0=10^8$~cm ($r_0=10^9$~cm) is adopted.
Here, the luminosity of the initial fireball $L_{\rm w}$ is estimated as $L_{\rm w}\sim E_{\rm k,0}/2.5$~s.
Corresponding, the related photosphere emission is plotted in the right panel of Figure~{\MyFigG},
where the observed power-law radiation spectrum in the period of $t_{\rm obs}\sim[0,5]$~s
is shown with a black solid.

\acknowledgments
We thank the anonymous referee of this work for useful comments and suggestions that improved the paper.
We acknowledge the use of the Fermi archive's public data.
We appreciate Xing Yang for his help in this work.
This work is supported by the National
Natural Science Foundation of China (grant Nos. 12273005,
11673006, U1938116, U1938201, U1731239, and U1938106),
the Guangxi Science Foundation (grant Nos. 2018GXNSFFA281010,
2017AD22006, 2018GXNSFGA281007, and 2018GXNSFDA281033), and China Manned Spaced Project (CMS-CSST-2021-B11).

%%%%%%%%%%%%%%%%%%%%%%%%%%%%%%%%%%%%%%%%%%%%%%%%%%%%%%%%%%%%%%%%%%%%%%%%%%%%%%%%%%%%%%%%%%%%%%%%%%%
%%%%%%%%%%%%%%%%%%%%%%%%%%%%%%%%%%%%%%%%%%%%%%%%%%%%%%%%%%%%%%%%%%%%%%%%%%%%%%%%%%%%%%%%%%%%%%%%%%%

%%%%%%%%%%%%%%%%%%%%%%%%%%%%%%%%%%%%%%%%%%%%%%%%%%%%%%%%%%%%%%%0%%%%%%%%%0%%%%%%%%0%%%%%%%%%%%%%%%%%%%%%%%%%%%%%%%%%%%%%%%%%%%%%%%%%%%%%%%%%%%%%%%%%%%%%%%%%%
\clearpage

%%%%%%%-----------------------------------------------
\begin{figure}
\centering
\begin{tabular}{cc}
\includegraphics[angle=0,scale=0.28, trim=200 0 0 0]{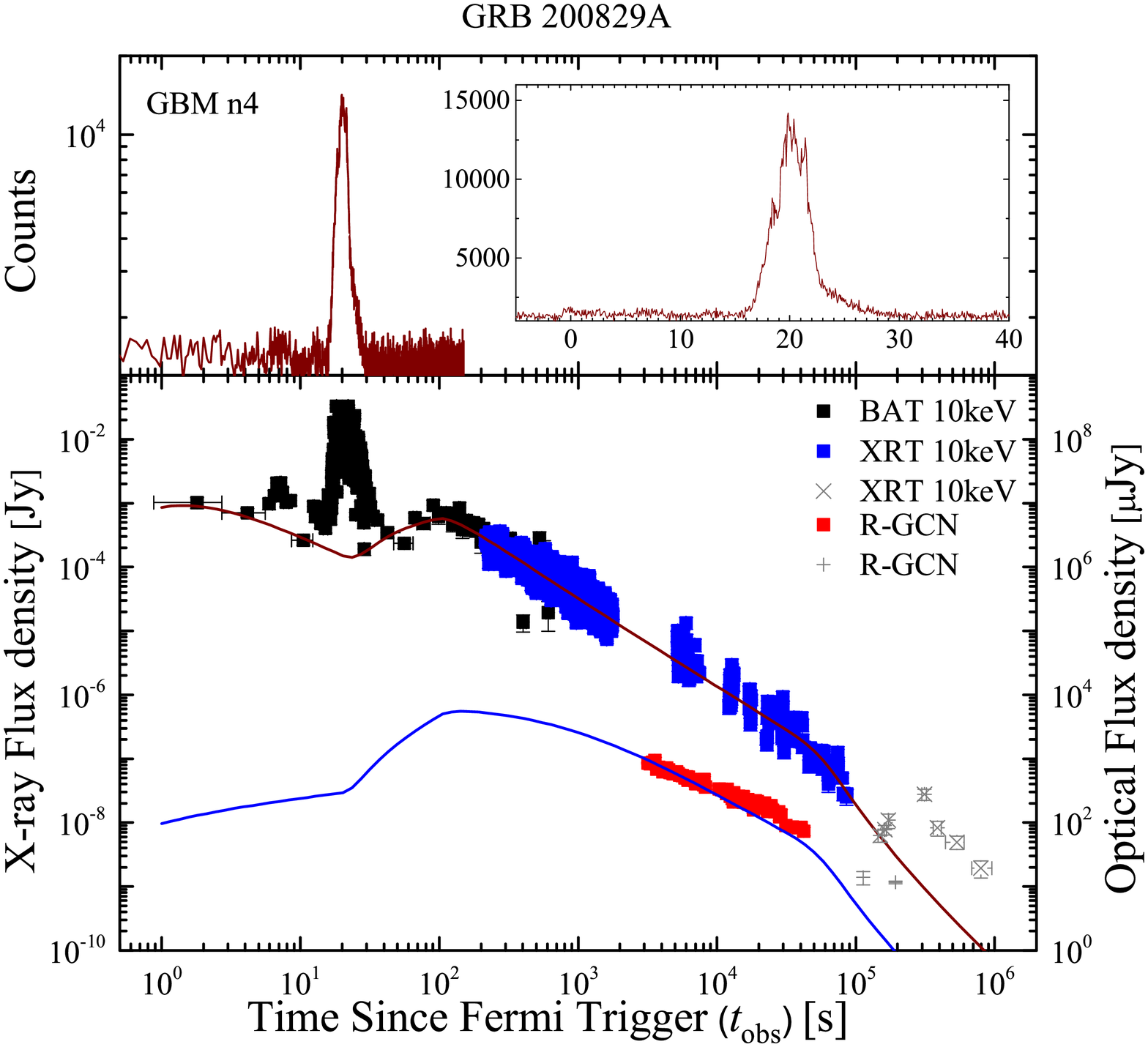}&
\includegraphics[angle=0,scale=0.40, trim=0 175 100 0]{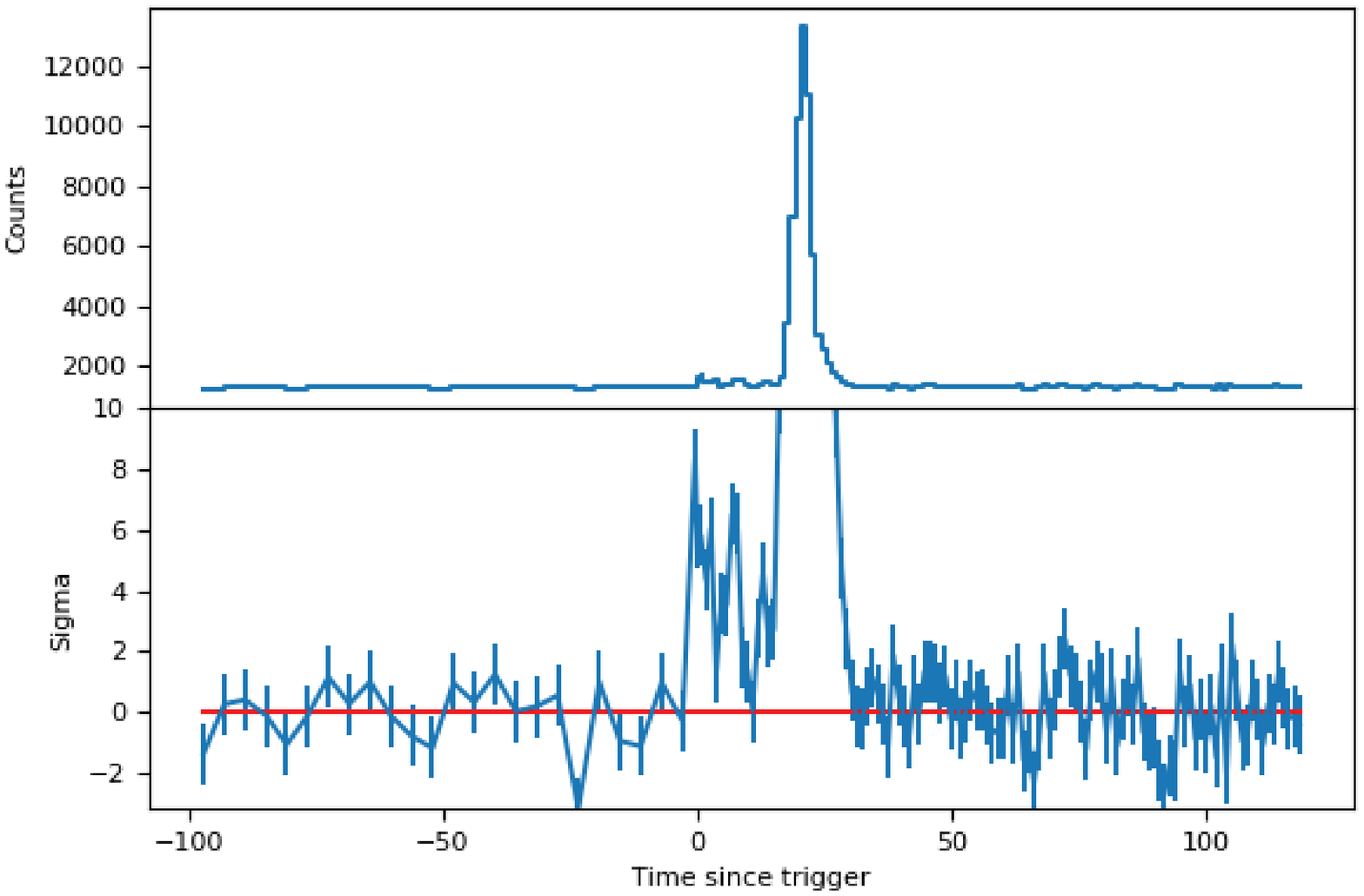}\\
\end{tabular}
\caption{{\em Left panel---}light-curves of GRB~200829A from prompt emission to its afterglows
and the BAT/XRT data are the flux density at 10~keV extrapolated from BAT/XRT observation,
where the inset of the upper-right panel shows the prompt $\gamma$-rays in the linear spaces.
The MCMC fitting result based on the model in Appendix~\ref{AppendixB} is shown with wine line and blue line for
X-ray and optical data, respectively. Here, the data showed with gray  ``$\times$''  and ``$+$'' symbols are not used in our fittings.
{\em Right panel---} GBM light-curve of GRB~200829A without background subtracted (upper panel)
and the signal significance (bottom panel),
where the background were estimated by fitting the light-curve before and after the burst with polynomial model.
It reveals that there is significantly photons in the period of $[0,10]$~s from GRB~200829A.
}\label{MyFigA}
\end{figure}

\begin{figure}
\begin{tabular}{ccc}
\includegraphics[angle=0,scale=0.23, trim=310 50 290 100]{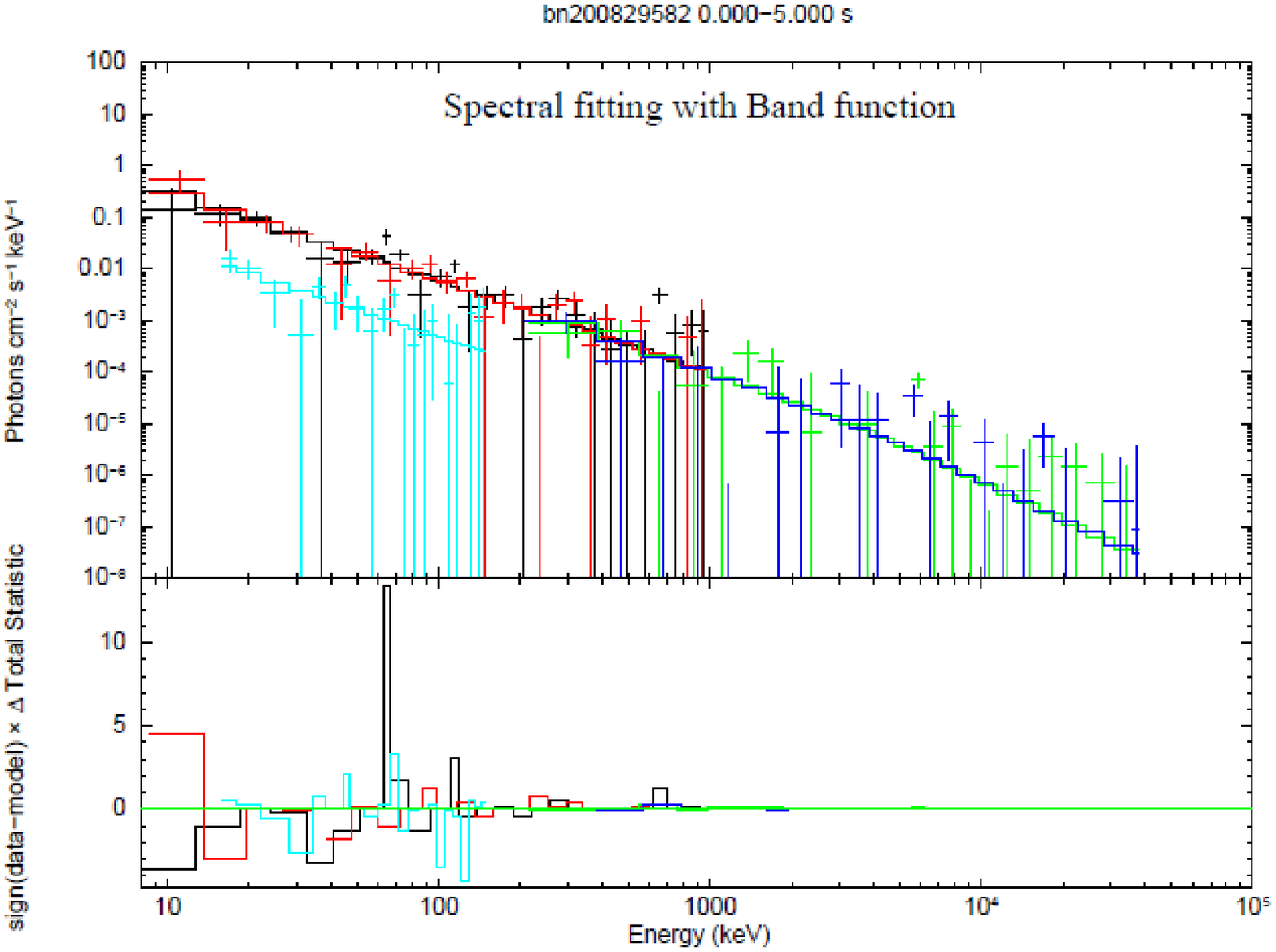} &
\includegraphics[angle=0,scale=0.23, trim=70 4 200 70]{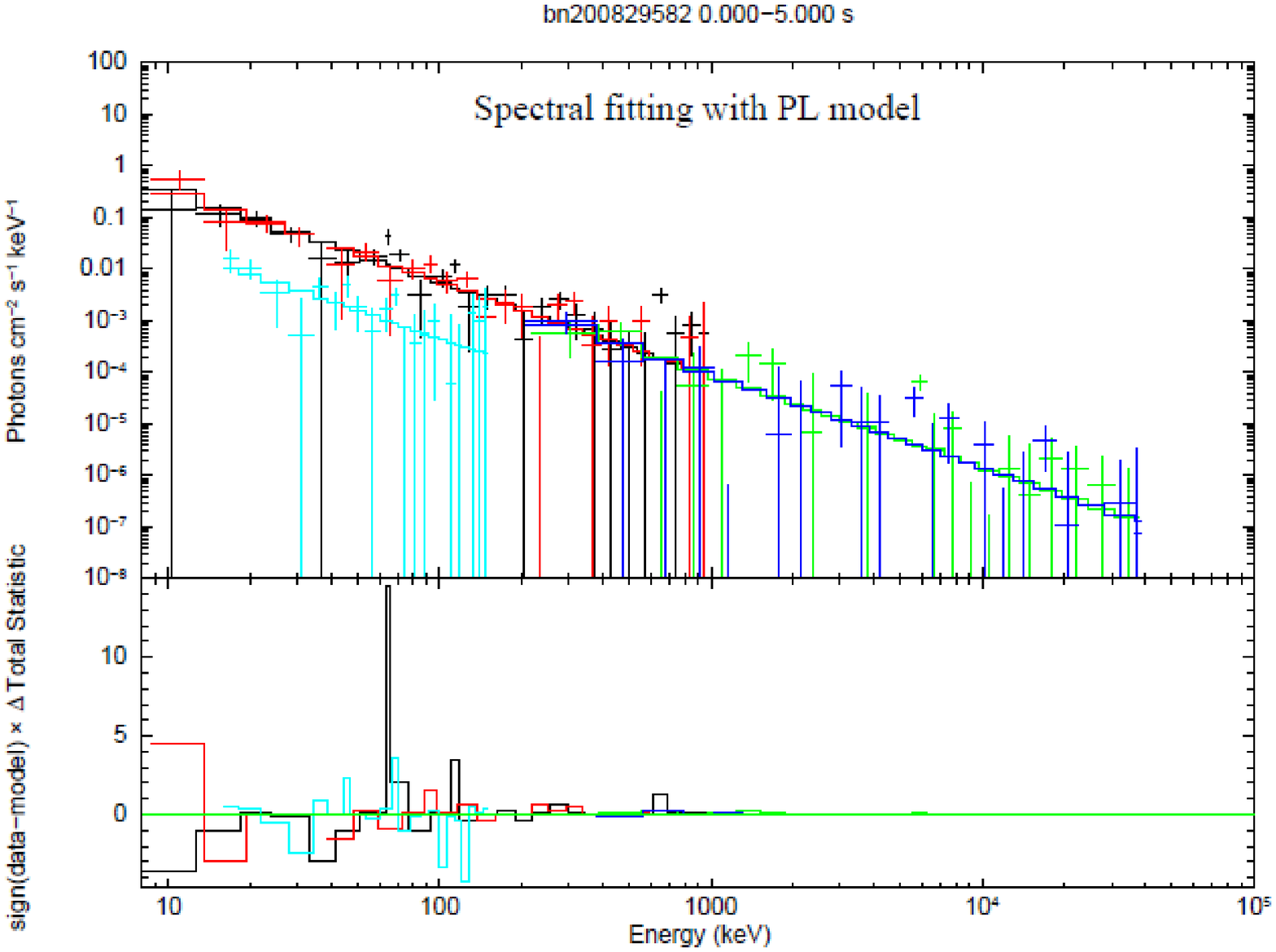} &
\includegraphics[angle=0,scale=0.27, trim=20 0 40 0]{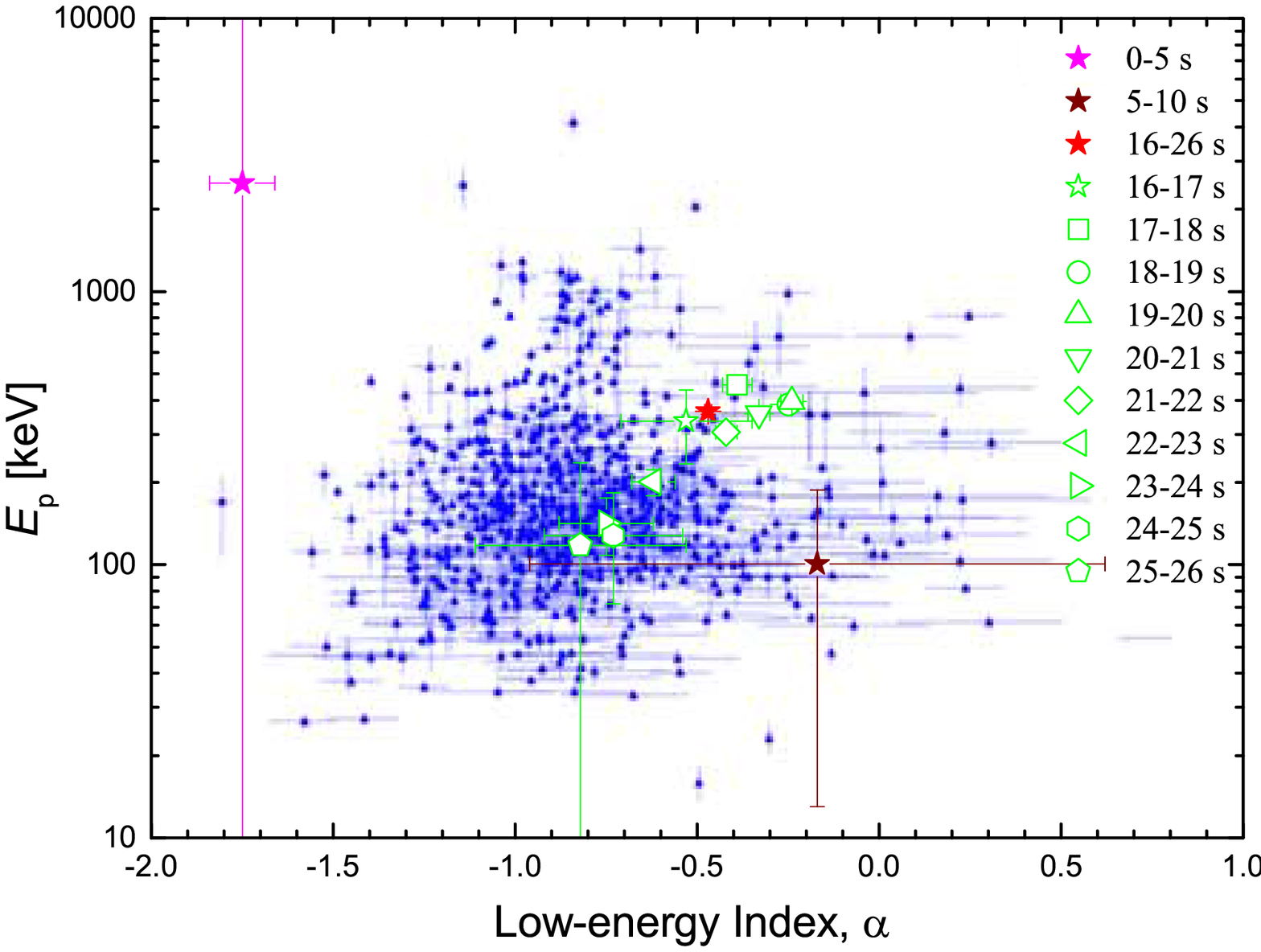} \\
\end{tabular}
\caption{Spectral fitting results of the very early prompt emission ($t_{\rm obs}\in[0,5]$~s) in GRB~200829A.
Here, the joint spectral fitting by combining \emph{Swift}-BAT and \emph{Fermi}-GBM observations
based on the Band function (left panel)
or power-law function (middle panel) are performed.
In addition, the relation of $E_{\rm p}$ and $\alpha$ based on the spectral fitting results with Band function
are plotted in right panel with ``$\bigstar$'' symbols,
where the blue symbols are from \cite{Poolakkil_S-2021-Preece_R-ApJ.913.60P}.
Here, the different green hollow symbols are the time-resolved spectral fitting results in the period of $[16,26]$~s.
}
\label{MyFigB}
\end{figure}

\clearpage

%%%%%%%%%%%%%%%%%%%%%%%%%%%%%%%%%%%%%%%%%%%%%%%%%%%%%%%%%%%%%%%%%%%%%%%%%%%%%%%%%%%%%%%%%%%%
\begin{figure}
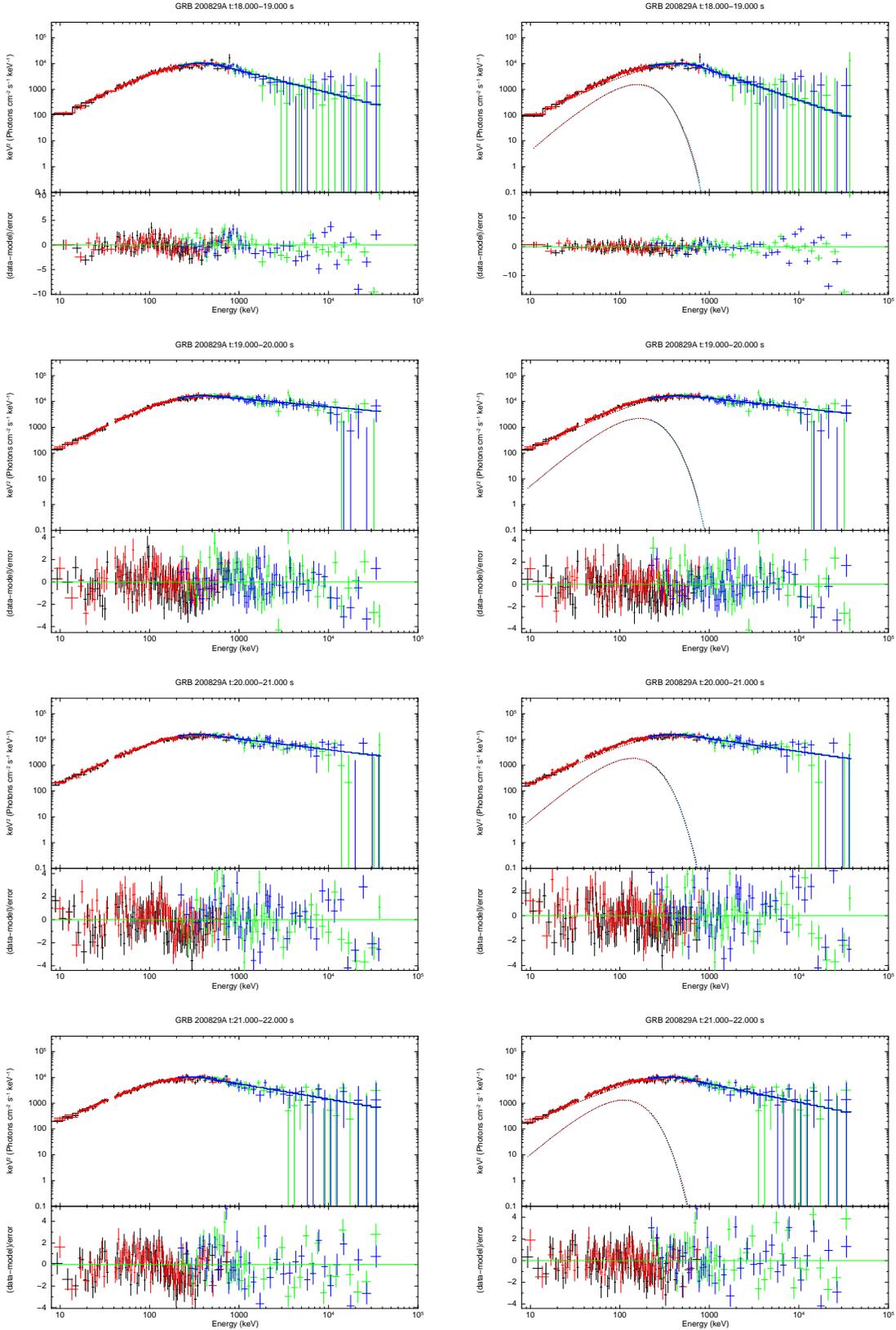

\begin{tabular}{cc}
\includegraphics[angle=-90,scale=0.28]{Fig_3_200829A_18.000-19.000.eps}&
\includegraphics[angle=-90,scale=0.28]{Fig_3_18.000-19.000_bband.eps}\\
\includegraphics[angle=-90,scale=0.28]{Fig_3_200829A_19.000-20.000.eps}&
\includegraphics[angle=-90,scale=0.28]{Fig_3_19.000-20.000_bband.eps}\\
\includegraphics[angle=-90,scale=0.28]{Fig_3_200829A_20.000-21.000.eps}&
\includegraphics[angle=-90,scale=0.28]{Fig_3_20.000-21.000_bband.eps}\\
\includegraphics[angle=-90,scale=0.28]{Fig_3_200829A_21.000-22.000.eps}&
\includegraphics[angle=-90,scale=0.28]{Fig_3_21.000-22.000_bband.eps}\\
\end{tabular}
\caption{Spectral fitting results of the bright spiky $\gamma$-ray pulse in the period of $t_{\rm obs}\in[18,22]$~s based on
Band function (left panel) or Band+BB model (right panel).}
\label{MyFigC}
\end{figure}
%%%%%%%%%%%%%%%%%%%%%%%%%%%%%%%%%%%%%%%%%%%%%%%%%%%%%%%%%%%%%%%%%%%%%%%%%%%%%%%%%%%%%%%%%%%%
%%%%%%%%%%%%%%%%%%%%%%%%%%%%%%%%%%%%%%%%%%%%%%%%%%%%%%%%%%%%%%%%%%%%%%%%%%%%%%%%%%%%%%%%%%%%

\begin{figure}
\begin{tabular}{ccc}
\includegraphics[angle=0,scale=0.21,trim=180 0 100 0]{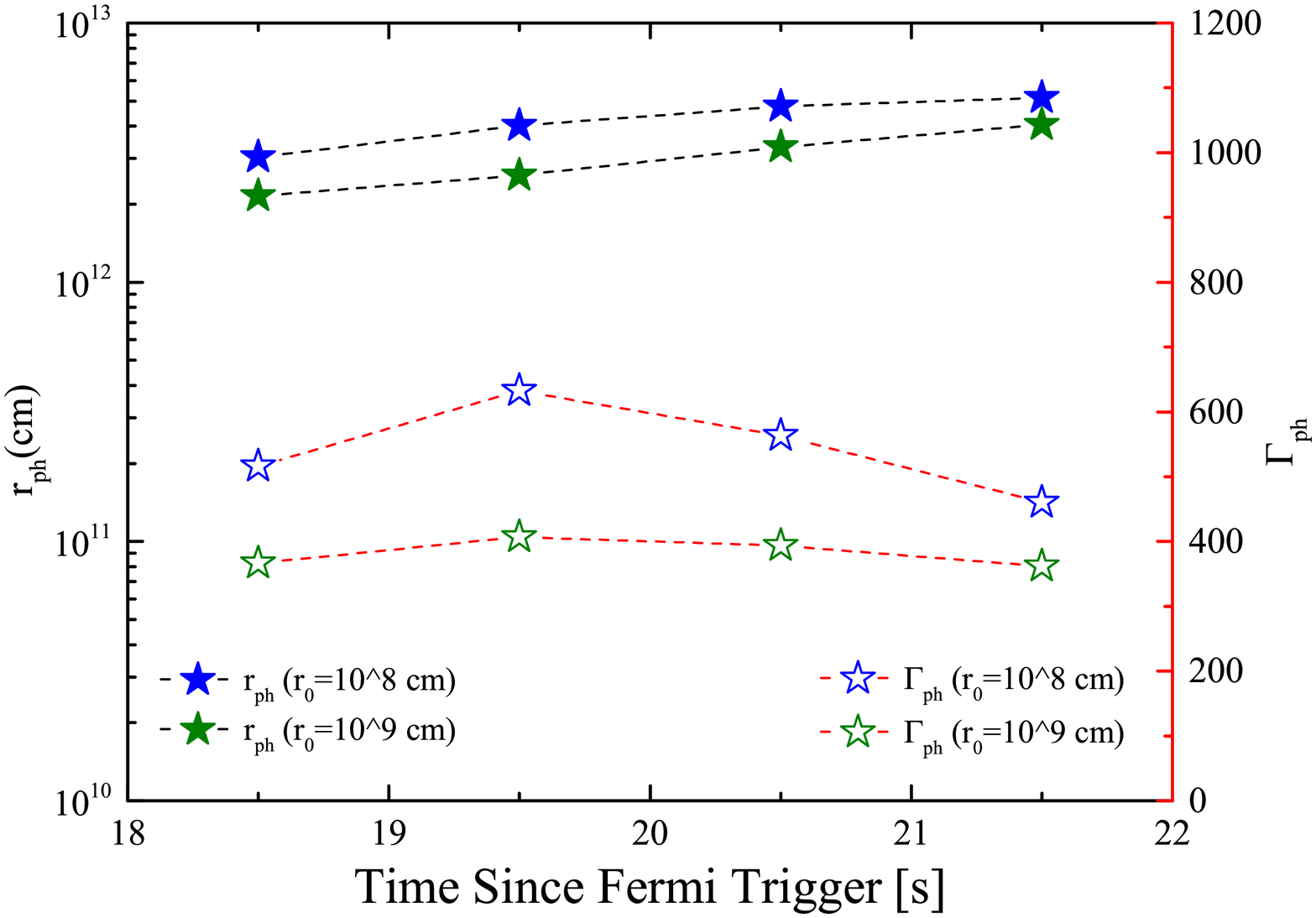}&
\includegraphics[angle=0,scale=0.21,trim=0 0 100 0]{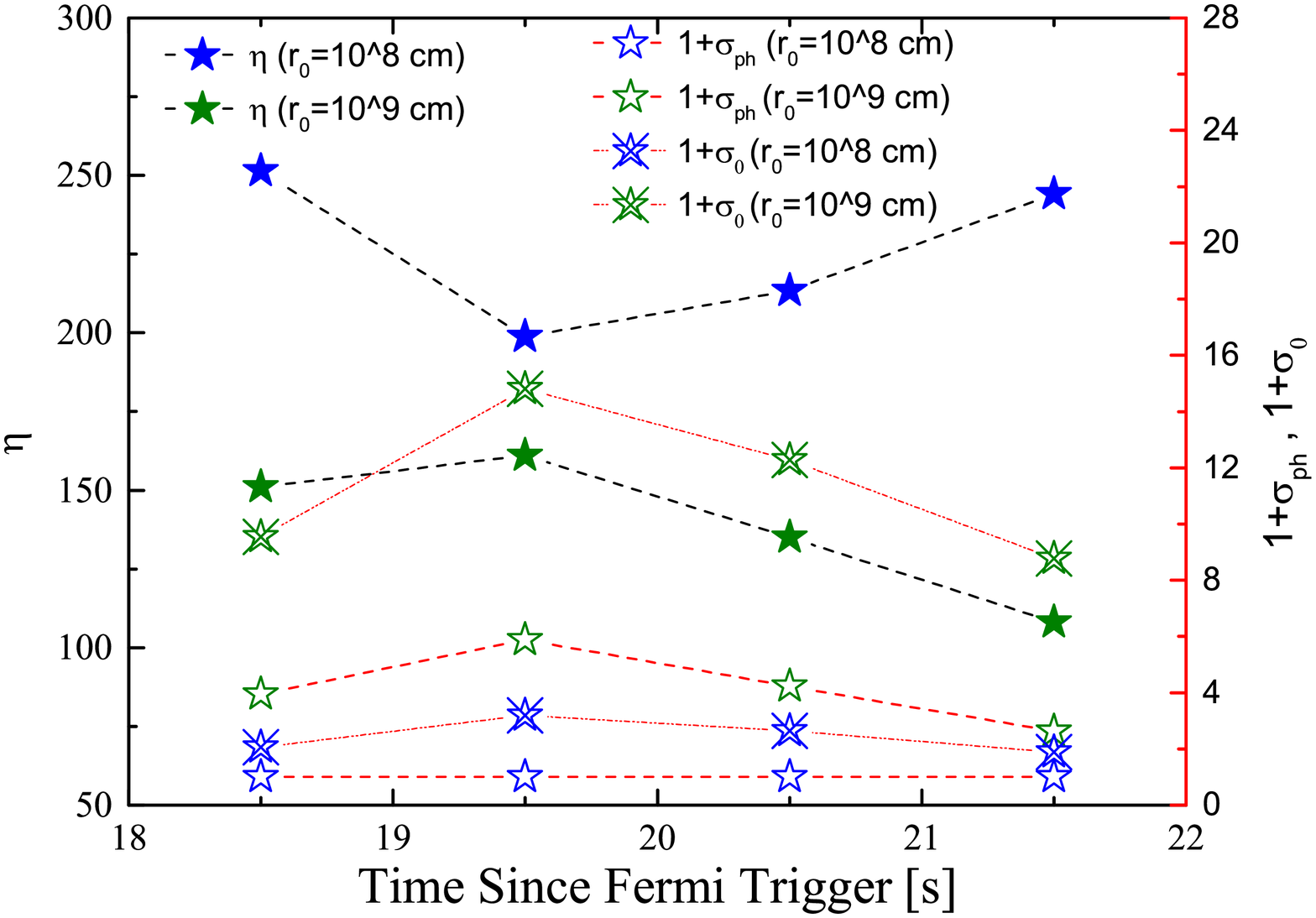}&
\includegraphics[angle=0,scale=0.21]{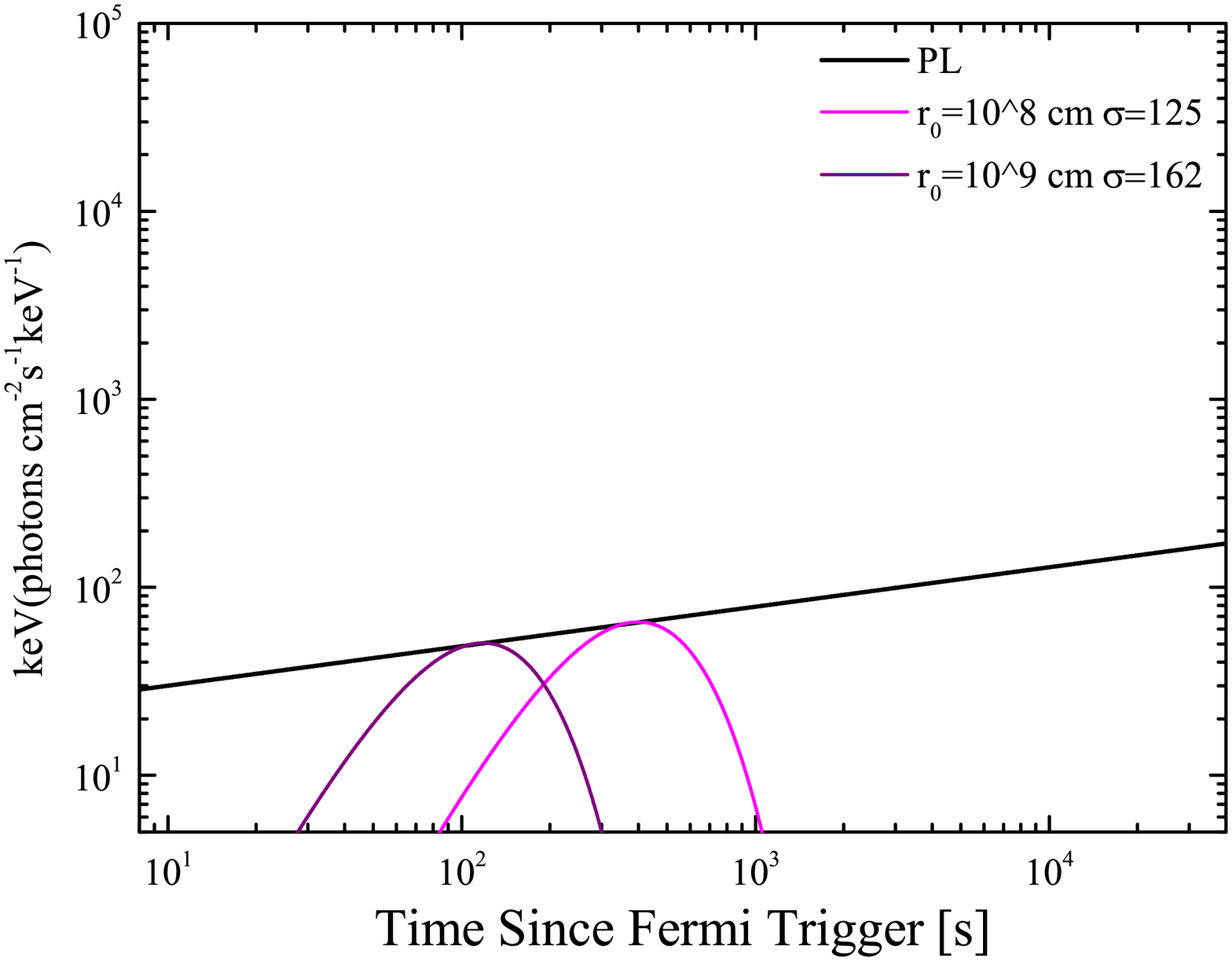}\\
\end{tabular}
\caption{{\em Left and middle panels---}Temporal evolution of derived properties
($r_{\rm ph}$, $\Gamma_{\rm ph}$, $\eta$, $1+\sigma_{\rm ph}$, and $1+\sigma_0$)
based on the blackbody radiation component found in the bright spiky $\gamma$-ray pulse.
{\em Right panel---}Power-law radiation spectrum found in the period of $t_{\rm obs}\in[0,5]$~s (solid line)
and the predicted lower limits of the photospheric emission (magenta and purple solid lines) for different parameters.
}
\label{MyFigD}
\end{figure}
%%%%%%%%%%%%%%%%%%%%%%%%%%%%%%%%%%%%%%%%%%%%%%%%%%%%%%%%%%%%%%%%%%%%%%%%%%%%%%%%%%%%%%%%%%%%
%%%%%%%%%%%%%%%%%%%%%%%%%%%%%%%%%%%%%%%%%%%%%%%%%%%%%%%%%%%%%%%%%%%%%%%%%%%%%%%%%%%%%%%%%%%%

\clearpage
\begin{figure}
\centering
\includegraphics[angle=0,scale=0.35]{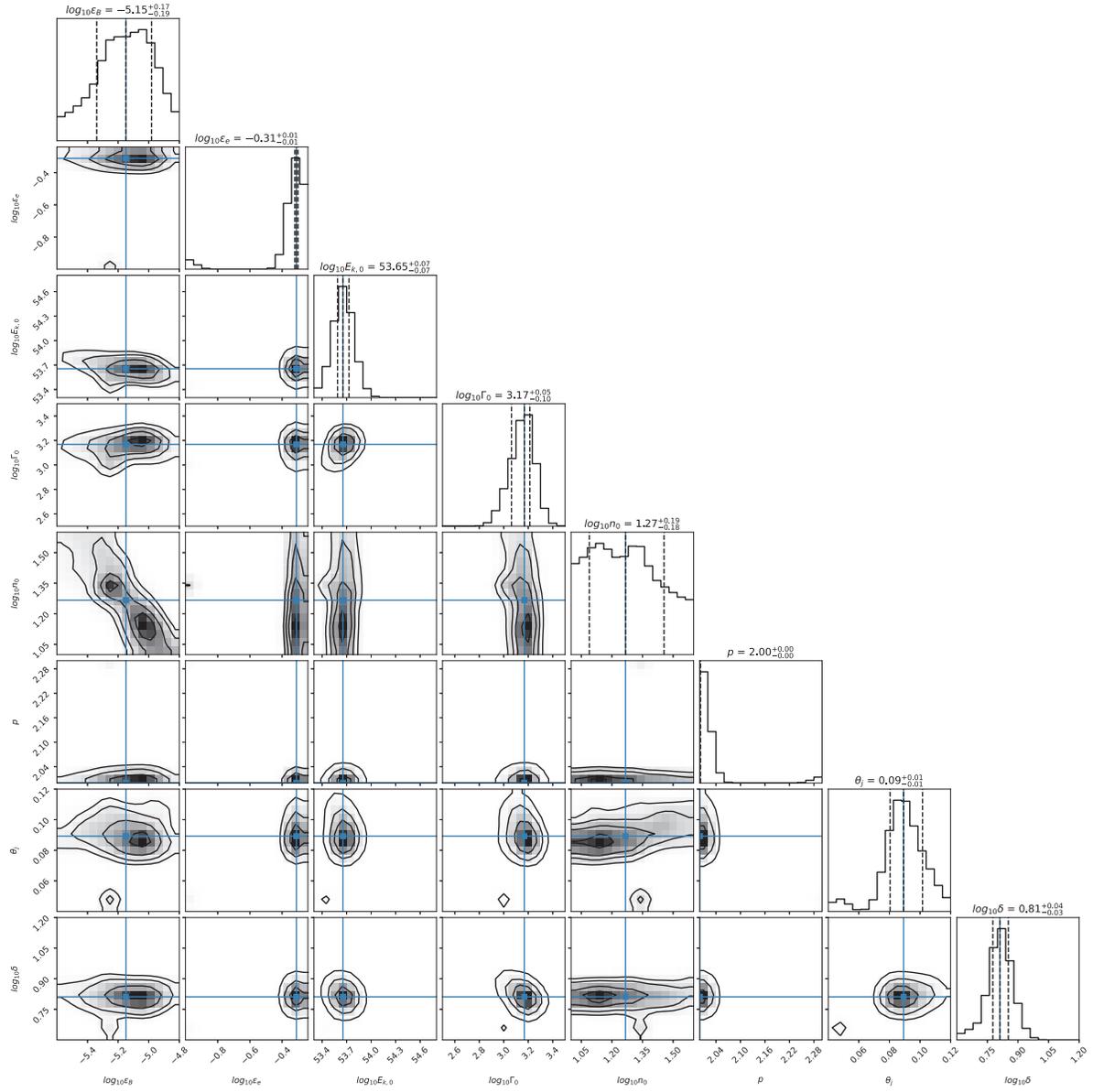} \\
\caption{Posterior probability density functions for the physical parameters of the external-forward shock
in GRB~200829A from MCMC simulations.}
\label{MyFigG}
\end{figure}
%%%%%%%%%%%%%%%%%%%%%%%%%%%%%%%%%%%%%%%%%%%%%%%%%%%%%%%%%%%%%%%%%%%%%%%%%%%%%%%%%%%%%%%%%%%%
\clearpage
\begin{table}
\tablewidth{500pt}
\caption{Spectral fitting results of the very early prompt emission in GRB~200829A.}
\centering{
\begin{tabular}{ccccccc}
\hline\hline
  Time interval (s) & Model  & $\alpha$ (or $\hat{\Gamma}$) \tablenotemark{a} & $\beta$ &$E_0  ({\rm keV})$ & $N_0$\tablenotemark{b} &$\chi_r^{2}$   \\
\hline
$[0, 5]$&PL  &  $-1.79\pm0.06$  &  - &  - &  $21.59\pm5.98$ & $1.08$ \\
\hline
$[0, 5]$&Band  &  $-1.75\pm0.09$ &  $-2.42\pm5.08$&  $9976.67\pm51113.36$&  $0.006\pm0.0006$ & $1.08$  \\
$[5, 10]$&Band  &  $-0.17\pm0.79$ &  $-2.25\pm0.27$&  $54.94\pm41.55$&  $0.05\pm0.06$ & $0.99$  \\
\hline
\end{tabular}}
\tablenotetext{a}{The photon spectral index $\hat{\Gamma}$ is for PL model and $\alpha$ is for Band function model.}
\tablenotetext{b}{$N_0$ is in unit of ${\rm photons \cdot cm^{-2} \cdot s^{-1}\cdot keV^{-1}}$.}
\label{MyTabA}
\end{table}

%%%%%%%%%%%%%%%%%%%%%%%%%%%%%%%%%%%%%%%%%%%%%%%%%%%%%%%%%%%%%%%%%%%%%%%%%%%%%%%%%%%%%%%%%%%%
\clearpage
\begin{table}
\tablewidth{0pt}
\caption{Spectral fitting results of the bright spicky $\gamma$-ray pulse in GRB~200829A.}
\centering
\resizebox{\linewidth}{18mm}{
\begin{tabular}{c|ccccc|ccccccc|c}
\hline\hline
Time interval (s) &$\quad$ & Band &&$\quad$ & &&$\quad$&Band + BB&$\quad$&&$\quad$& &\\
& $\alpha$ & $E_0$(keV) & $\beta$ &$N_{\rm 0}$\tablenotemark{a}& BIC & $\alpha$ & $E_0$(keV) & $\beta$ &$N_{\rm 0}$\tablenotemark{a}& kT(keV) &$K$\tablenotemark{a}& BIC &$\Delta$BIC\tablenotemark{b}\\
\hline
$[16, 26]$	&	-0.47 	$\pm$	0.01 	&	231.41 	$\pm$	4.11 	&	-2.47 	$\pm$	0.02 	&	0.41 	$\pm$	0.00 	&	948.83 	&	-0.52 	$\pm$	0.02 	&	286.22 	$\pm$	9.53 	&	-2.56 	$\pm$	0.02 	&	0.32 	$\pm$	0.01 	&	32.82 	$\pm$	1.68 	&	16.34 	$\pm$	1.74 	&	841.76 	&	107.07 	\\
$[16, 17]$	&	-0.53 	$\pm$	0.18 	&	225.71 	$\pm$	60.30 	&	-2.29 	$\pm$	0.19 	&	0.06 	$\pm$	0.01 	&	510.41 	&	-0.80 	$\pm$	0.20 	&	599.36 	$\pm$	420.69 	&	-3.40 	$\pm$	2.15 	&	0.02 	$\pm$	0.00 	&	35.86 	$\pm$	6.62 	&	7.19 	$\pm$	1.95 	&	517.64 	&	-7.23 	\\
$[17, 18]$	&	-0.40 	$\pm$	0.04 	&	283.01 	$\pm$	16.61 	&	-2.83 	$\pm$	0.13 	&	0.23 	$\pm$	0.01 	&	550.50 	&	-0.44 	$\pm$	0.07 	&	350.38 	$\pm$	34.98 	&	-3.15 	$\pm$	0.24 	&	0.18 	$\pm$	0.01 	&	42.30 	$\pm$	5.88 	&	17.22 	$\pm$	5.21 	&	548.53 	&	1.96 	\\
$[18, 19]$	&	-0.25 	$\pm$	0.03 	&	216.73 	$\pm$	7.72 	&	-2.86 	$\pm$	0.07 	&	0.58 	$\pm$	0.01 	&	595.01 	&	-0.31 	$\pm$	0.05 	&	273.54 	$\pm$	17.43 	&	-3.16 	$\pm$	0.12 	&	0.41 	$\pm$	0.02 	&	40.05 	$\pm$	3.53 	&	39.41 	$\pm$	7.56 	&	572.53 	&	22.49 	\\
$[19, 20]$	&	-0.25 	$\pm$	0.03 	&	221.12 	$\pm$	7.34 	&	-2.32 	$\pm$	0.02 	&	0.92 	$\pm$	0.02 	&	536.65 	&	-0.36 	$\pm$	0.05 	&	297.34 	$\pm$	26.24 	&	-2.38 	$\pm$	0.03 	&	0.65 	$\pm$	0.05 	&	42.56 	$\pm$	3.85 	&	57.77 	$\pm$	14.49 	&	523.84 	&	12.81 	\\
$[20, 21]$	&	-0.32 	$\pm$	0.03 	&	207.24 	$\pm$	6.69 	&	-2.42 	$\pm$	0.02 	&	1.04 	$\pm$	0.02 	&	658.42 	&	-0.40 	$\pm$	0.05 	&	264.04 	$\pm$	18.63 	&	-2.50 	$\pm$	0.03 	&	0.76 	$\pm$	0.05 	&	35.26 	$\pm$	3.28 	&	47.93 	$\pm$	10.16 	&	636.37 	&	22.05 	\\
$[21, 22]$	&	-0.41 	$\pm$	0.03 	&	188.50 	$\pm$	7.07 	&	-2.59 	$\pm$	0.04 	&	0.90 	$\pm$	0.03 	&	601.88 	&	-0.45 	$\pm$	0.05 	&	232.22 	$\pm$	15.55 	&	-2.71 	$\pm$	0.06 	&	0.66 	$\pm$	0.04 	&	27.89 	$\pm$	2.56 	&	33.86 	$\pm$	5.88 	&	576.20 	&	25.67 	\\
$[22, 23]$	&	-0.60 	$\pm$	0.06 	&	139.79 	$\pm$	13.33 	&	-2.31 	$\pm$	0.05 	&	0.41 	$\pm$	0.03 	&	497.37 	&	-0.84 	$\pm$	0.11 	&	254.12 	$\pm$	56.51 	&	-2.46 	$\pm$	0.11 	&	0.22 	$\pm$	0.04 	&	24.06 	$\pm$	3.31 	&	13.07 	$\pm$	4.12 	&	501.53 	&	-4.17 	\\
$[23, 24]$	&	-1.03 	$\pm$	0.09 	&	195.52 	$\pm$	37.43 	&	-2.45 	$\pm$	0.17 	&	0.14 	$\pm$	0.02 	&	490.10 	&	-1.20 	$\pm$	0.18 	&	340.63 	$\pm$	181.20 	&	-2.48 	$\pm$	0.31 	&	0.18 	$\pm$	0.23 	&	23.79 	$\pm$	3.84 	&	1.04 	$\pm$	9.77 	&	495.21 	&	-5.11 	\\
$[24, 25]$	&	-0.59 	$\pm$	0.23 	&	80.59 	$\pm$	24.47 	&	-2.23 	$\pm$	0.10 	&	0.22 	$\pm$	0.08 	&	547.41 	&	-1.35 	$\pm$	0.15 	&	582.80 	$\pm$	366.04 	&	-2.50 	$\pm$	1.23 	&	0.04 	$\pm$	0.01 	&	22.00 	$\pm$	2.69 	&	7.76 	$\pm$	1.46 	&	559.13 	&	-11.72 	\\
$[25, 26]$	&	-0.86 	$\pm$	0.30 	&	102.82 	$\pm$	56.82 	&	-2.19 	$\pm$	0.16 	&	0.09 	$\pm$	0.05 	&	513.84 	&	-1.19 	$\pm$	1.09 	&	203.83 	$\pm$	657.78 	&	-2.13 	$\pm$	0.24 	&	0.07 	$\pm$	0.21 	&	21.75 	$\pm$	10.54 	&	0.71 	$\pm$	4.79 	&	526.66 	&	-12.82 	\\
\hline\hline
\end{tabular}}
\tablenotetext{a}{$N_0$ is in unit of ${\rm photons \cdot cm^{-2} \cdot s^{-1}\cdot keV^{-1}}$;
\emph{K} is the \emph{L$_{39}$}/ \emph{D$_{10}^2$}, where \emph{L$_{39}$} is the
	source luminosity in units of {10$^{39}$} erg/s and \emph{D$_{10}$}
is the distance to the source in units of 10 kpc.}
\tablenotetext{b}{The $\Delta$BIC is the value of $\rm BIC_{\rm Band}-BIC_{\rm Band+BB}$.}
\label{MyTabB}
\end{table}
%%%%%%%%%%%%%%%%%%%%%%%%%%%%%%%%%%%%%%%%%%%%%%%%%%%%%%%%%%%%%%%%%%%%%%%%%%%%%%%%%%%%%%%%%%%%
\clearpage
\begin{table}
\tablewidth{500pt}
\caption{Results of spectral fits for $t_{\rm obs}\in[230,52000]$~s of GRB~200829A. }
\centering{
\begin{tabular}{clllc}
\hline\hline
  GRB   &Interval(s)& Band & $\chi_r^{2}$ &  $\hat{\Gamma}$   \\
\hline
GRB~200829A   &  230-700& BAT+XRT &$1.00$   &  $-2.05\pm0.04$    \\
     &  700-2000& XRT &$1.11$   &  $-1.75\pm0.01$    \\
    &    5116-7428& XRT &  $0.94$  &  $-1.76\pm0.05$   \\
     &  12119-13162& XRT &$1.09$   &  $-1.83\pm0.06$    \\
     &  28067-52000& XRT & $1.19$  &  $-1.89\pm0.06$    \\
\hline
\end{tabular}}
\label{MyTabC}
\end{table}

%%%%%%%%%%%%%%%%%%%%%%%%%%%%%%%%%%%%%%0%%%%0%%%%%%%%%%%%%%%%%%%%%%%%%%%%%%%%%%%%%%%
%%%%%%%%%%%%%%%%%%%%%%%%%%%%%%%%%%%%%%%%%%%%%%%%%%%%%%%%%%%%%%%%%%%%%%%%%%%%%%%%%%
\clearpage
\appendix
\section{Discussion about the prompt emission of GRB~200829A in the period of $[0, 5]$~s}\label{AppendixA}
In this section, we present a comprehensive discussion about the radiation spectrum in the prompt emission of GRB~200829A in the period of $[0, 5]$~s.
We would like to conclude that
the intrinsic radiation spectrum in this period
may be consistent with a PL spectral model with $\hat{\Gamma} \sim -1.7$ or
a Band function with a break at $\sim10$~MeV and power-law index $\sim -1.7$
in its low-energy regime ($E\lesssim 10$~MeV),
rather than a Band function with $\alpha\sim -1$, $\beta\sim -3$, and $E_{\rm p}\sim 200$~keV.
This conclusion is made based on the comprehensive comparison between the spectral fitting results
on the observational data and those on the synthetic data of \emph{Fermi} observation.
Here, the synthetic data of \emph{Fermi} observation is generated
based on the python source package {\tt threeML}\footnote{https://github.com/threeML/threeML} (\citealp{Vianello_G-2015-Lauer_RJ-arXiv150708343V})
and the Band function with $\alpha=-1$, $\beta=-3$, and $E_{\rm p}=200$~keV is adopted as
the intrinsic radiation spectrum to produce synthetic data.
In addition, the signal significance of the synthetic data
is set as that of the observational data of GRB~200829A in the period of $[0, 5]$~s.
The spectral fittings in this section are performed based on the MCMC method
to produce posterior predictions for the model parameters\footnote{This method is different from
that used in the main text of the present paper.
In the main text, the spectral model parameters are obtained based on the package {\tt Xspec} by maximizing the likelihood.
However, one can find that the model parameters are consistent with each other in these two fitting methods.}
and the python source package {\tt emcee} (\citealp{Foreman-Mackey_D-2013-Hogg_DW-PASP.125.306F})
is used for our MCMC sampling. The spectral fitting results are reported in Table~{\MyTabD}.

The reasons for our above conclusion are as follows.
\begin{enumerate}
\item
In the spectral fitting, the values of
``Residuals ($\sigma$)'' (see the bottom part in each panel of Figure~{\MyFigE}) provides
the most important information to confront the spectral model with the observed data.
A good spectral model for the observational data should provide a well distribution of ``Residuals ($\sigma$)'',
such as that shown in the bottom part of the upper-right panel in Figure~{\MyFigE}.
In Figure~{\MyFigE}, the upper-left and upper-right panels show the spectral fitting results on
the synthetic data with a PL model and a Band function, respectively.
Since the intrinsic radiation spectrum of the synthetic data is a Band function with $E_{\rm p}=200$~keV,
the spectral fitting on the synthetic data with a Band function should provide an optimal fitting.
Actually, the values of the corresponding ``Residuals ($\sigma$)'' are indeed well distributed around zero.
In the spectral fitting on the synthetic data with a PL model,
however, the values of ``Residuals ($\sigma$)'' appear as positive around $E_{\rm p}$ and negative below/above $\sim E_{\rm p}$.
It reveals that even though the Band function with $\alpha=-1$, $\beta=-3$, and $E_{\rm p}=200$~keV
can be described as a PL model with $\hat{\Gamma}\sim -1.65$ (see second line of Table~{\MyTabD}),
the observational data would exceed the PL model around $E_{\rm p}$
and fail to reach the PL model below/above $\sim E_{\rm p}$.
This behavior is consistent with the theoretical expectation.

$\;\;$ In the bottom-left and bottom-right panels of Figure~{\MyFigE},
we show the spectral fitting results on the observational data of GRB~200829A in the period of $[0, 5]$~s
with a PL spectral model and a Band function, respectively.
The spectral fitting results are also reported in the fourth and fifth lines of Table~{\MyTabD}.
One can find that ``Residuals ($\sigma$)'' in these two panels are well distributed around zero,
which is very similar to that in the upper-right panel.
It implies that the intrinsic radiation spectrum of this period should be
consistent with a PL spectral model with $\hat{\Gamma} \sim -1.7$ or
a Band function with a break at $\sim10$~MeV and power-law index $\sim -1.7$
in its low-energy regime ($E\lesssim 10$~MeV),
rather than a Band function with $\alpha\sim -1$, $\beta\sim -3$, and $E_{\rm p}\sim 200$~keV.
This is because that if the intrinsic radiation spectrum of the observational data is a Band function with $\alpha\sim -1$, $\beta\sim -3$, and $E_{\rm p}=200$~keV,
the values of ``Residuals ($\sigma$)'' would be positive $\sim 200$~keV
and negative below/above $\sim 200$~keV on average.
However, this behavior could not be evidently found in the bottom-left panel of Figure~{\MyFigE}.

\item
If the intrinsic radiation spectrum in this period is the Band function with $E_0\sim 200$~keV,
the spectral fitting results on the low-energy regime, e.g., the energy band of \emph{Swift}-BAT (15-150~keV),
with a PL spectral model would be very different from that on the energy band of \emph{Fermi}-GBM instrument (8~keV-40~MeV).
Then, we perform the spectral fittings on the data in the 15-150~keV energy band.
The posterior probability density functions for the physical parameters of the spectral model are shown in Figure~{\MyFigF},
where the upper and bottom panels are the spectral fitting results on
the synthetic data and the observational data in the 15-150~keV energy band, respectively.
A PL spectral model and Band function are adopted in the spectral fittings for the left and right panels, respectively.
It is shown that the spectral fittings on the synthetic data with a PL spectral model for different energy regime
are indeed presented very different values of power-law index $\hat{\Gamma}$, i.e., $\hat{\Gamma}=-1.65_{-0.04}^{+0.04}$
for the 8~keV-40~MeV energy band and $\hat{\Gamma}=-1.44_{-0.10}^{+0.10}$ for the 15-150~keV energy band.
Interestingly, the spectral fittings on the synthetic data with a Band function
almost report the same values of $\alpha$, $\beta$, and $E_0$
for the 15-150~keV energy band and the 8~keV-40~MeV energy band.
According to the fitting results reported in Table~{\MyTabD},
one can find that the spectral fittings on the observational data in the 15-150~keV energy band
and those in the 8~keV-40~MeV energy band are almost presented the same fitting results.
Please comparing the eighth line with the fourth line, or the ninth line with the fifth line in Table~{\MyTabD}.
It implies that the radiation spectrum in this period should be consistent with
a PL spectral model with $\hat{\Gamma} \sim -1.7$ or
a Band function with a break at $\sim10$~MeV and power-law index $\sim -1.7$
in its low-energy regime ($E\lesssim 10$~MeV),
rather than a Band function with $\alpha\sim -1$, $\beta\sim -3$, and $E_{\rm p}\sim 200$~keV.
\end{enumerate}

In summary, by comparing the spectral fitting results on the observational data
to those on the synthetic data, we can conclude that
the intrinsic radiation spectrum in this period should be consistent with
a PL spectral model with $\hat{\Gamma} \sim -1.7$ or
a Band function with a break at $\sim10$~MeV and power-law index $\sim -1.7$
in its low-energy regime ($E\lesssim 10$~MeV).

\section{Model}\label{AppendixB}
In this section, the dynamics and the emission of the external-forward shock are presented as follows.
The dynamics of the external-forward shock can be described with the following equations
(e.g., \citealp{Sari_R-1998-Piran_T-ApJ.497L.17S, Huang_YF-1999-Dai_ZG-MNRAS.309.513H}):
\begin{equation}\label{Eq:Gamma}
\frac{{d\Gamma }}{{d{t_{{\rm{obs}}}}}} = \frac{1}{M'}\left[ {\frac{1}{{{c^2}}}\frac{{d{E_{{\rm{inj}}}}}}{{d{t_{{\rm{obs}}}}}} - ({\Gamma ^2} - 1)\frac{{dm}}{{d{t_{{\rm{obs}}}}}}} \right],
\end{equation}
\begin{equation}\label{Eq:mass}
\frac{{dm}}{dt_{\rm obs}} =4\pi \rho{R^2}\frac{{dR}}{{d{t_{{\rm{obs}}}}}} ,
\end{equation}
\begin{equation}
\frac{{dU'}}{{d{t_{{\mathop{\rm obs}\nolimits} }}}} =(1-\epsilon)(\Gamma  - 1){c^2}\frac{{dm}}{{d{t_{{\mathop{\rm obs}\nolimits} }}}},
\end{equation}
\begin{equation}\label{Eq:R}
\frac{{dR}}{{d{t_{{\rm{obs}}}}}} = \frac{{c\beta }}{{1 - \beta }}(1+z),
\end{equation}
\begin{equation}\label{Eq:_beta}
\beta  = \sqrt {1 - 1/{\Gamma ^2}} ,
\end{equation}
where $\Gamma$, ${d{E_{{\rm{inj}}}}/d{t_{{\rm{obs}}}}}$, $R$, $\epsilon$, and $c\beta$ are the Lorentz factor,
the energy injection rate (with respect to the observer time $t_{\rm obs}$), location, the radiation efficiency, and the velocity of the external-forward shock,
and $M'=M'_{\rm ej}+m+U'/c^2$ is the total mass, including the initial mass $M'_{\rm ej}=E_{\rm k,0}/[(\Gamma_0-1)c^2]$ of the ejecta,
the sweep-up mass $m$ from the circum-burst medium, and the internal energy $U'$ of the shocked material from the external shock.
Here, $E_{\rm k,0}$ is the initial isotropic kinetic energy of the fireball,
$\Gamma_0=\Gamma(t_{\rm obs}=0)$ is the initial bulk Lorentz factor of the fireball,
$c$ is the velocity of light, $z$ is the redshift of the burst, and $\rho$ is the density of the circum-burst environment.
Two cases of circum-burst medium, i.e., interstellar medium (ISM) and wind, are generally studied.
Correspondingly, we take (e.g., \citealp{Chevalier_RA-2000-Li_ZY-ApJ.536.195C})
\begin{equation}
\rho=
\left\{ {\begin{array}{*{20}{c}}
{5 \times 10^{11}A_*R^{-2}\,\rm g\cdot cm^{ - 1},}&{\rm{wind},}\\
{n_0m_p\,\rm cm^{ - 3},}&{\rm{ISM},}
\end{array}} \right.
\end{equation}
with $m_p$ being the proton mass, $A_*$ is a dimensionless constant.
For simplicity, the energy injection into the external shock due to the late activity of the central engine
is assumed with a constant energy injection rate over the period of $t_{\rm obs}\in[t_{\rm s}, t_{\rm e}]$,
where $t_{\rm s}$ and $t_{\rm e}$ are the beginning and the end of the energy injection, respectively.
By describing $E_{\rm inj}$ as $E_{\rm inj}=E_{\rm k,0}\delta$, one thus can have ${d{E_{{\rm{inj}}}}/d{t_{{\rm{obs}}}}}=E_{\rm k,0}\delta/(t_{\rm e}-t_{\rm s})$.

The main radiation mechanism of the external-forward shock in GRBs
is the synchrotron radiation of the sweep-up electrons
(\citealp{Sari_R-1998-Piran_T-ApJ.497L.17S, Sari_R-1999-Piran_T-ApJ.517L.109S}).
$\epsilon_{\rm e}$ and $\epsilon_{\rm B}$ are introduced to represent the fractions of the shock
energy used to accelerate electrons and contributing to the magnetic energy, respectively.
Then, the magnetic field behind the shock
is $B^{\prime}={(32\pi { \epsilon _B}\rho/m_{\rm p})^{1/2}}\Gamma c$.
The sweep-up electrons are accelerated
to a power-law distribution of Lorentz factor $\gamma_e$,
i.e., $Q\propto {\gamma '_e}^{ - p}$ for
$\gamma_{e, \min}^{\prime}
\leqslant \gamma_{e} \leqslant \gamma_{e,\max}^{\prime}$,
where $p (> 2)$ is the power-law index,
$\gamma_{e,\min}=\epsilon_e(p-2)m_{\rm p}\Gamma/[(p-1)m_{\rm e}]$ (\citealp{Sari_R-1998-Piran_T-ApJ.497L.17S}),
and $\gamma_{e,\max}=\sqrt {9m_{\rm{e}}^2{c^4}/(8B'q_e^3)}$ with $q_{\rm e}$ being the electron charge (e.g., \citealp{Kumar_P-2012-Hernandez_RA-MNRAS.427L.40K}).
Then, one can have $\epsilon=\epsilon_{\rm rad}\epsilon_e$
with $\epsilon_{\rm rad}={\rm min}\{1,(\gamma_{e,\min}/\gamma_{e,c})^{(p-2)}\}$
(\citealp{Fan_Y-2006-Piran_T-MNRAS.370L.24F}),
where $\gamma_{e,c}=6 \pi m_e c(1+z)/(\sigma_{\rm T}\Gamma {B^{\prime}}^2 t_{\rm obs})$
is the efficient cooling Lorentz factor of electrons.

Equations~(\ref{Eq:Gamma})-(\ref{Eq:_beta}) describe the evolution of hydrodynamic blastwave approximately.
A more rigorous treatment can be found in \cite{Nava_L-2013-Sironi_L-MNRAS.433.2107N} and \cite{Zhang_B-2018pgrb.book.Z} (see Eq.~(8.66) in this book).
For our studied burst,
the blastwave is affected by the energy injection
and thus its evolution could not be simply estimated with hydrodynamic equations in \cite{Nava_L-2013-Sironi_L-MNRAS.433.2107N} and \cite{Zhang_B-2018pgrb.book.Z}.
A more complicated equations are required.
For the phase without energy injection, we also present the light curve of afterglows
based on the hydrodynamic equations in \cite{Nava_L-2013-Sironi_L-MNRAS.433.2107N} and \cite{Zhang_B-2018pgrb.book.Z}.
It is found that the obtained light-curves of afterglows
are almost the same as those obtained with Equations~(\ref{Eq:Gamma})-(\ref{Eq:_beta}).

%0%%%%%-----------------------------------------------
\clearpage
\begin{figure}
\begin{tabular}{cc}
\includegraphics[angle=0,scale=0.50]{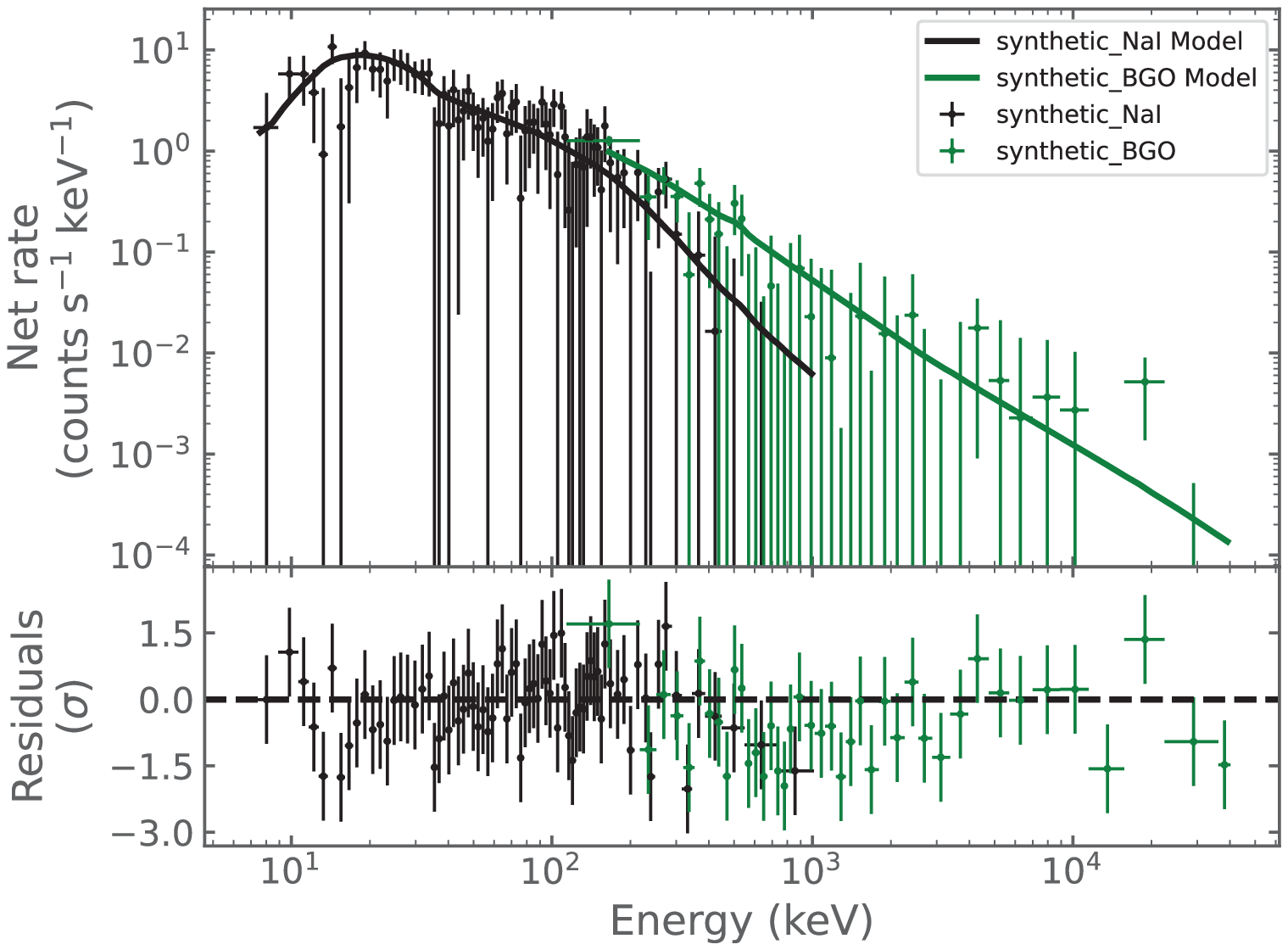}&
\includegraphics[angle=0,scale=0.50]{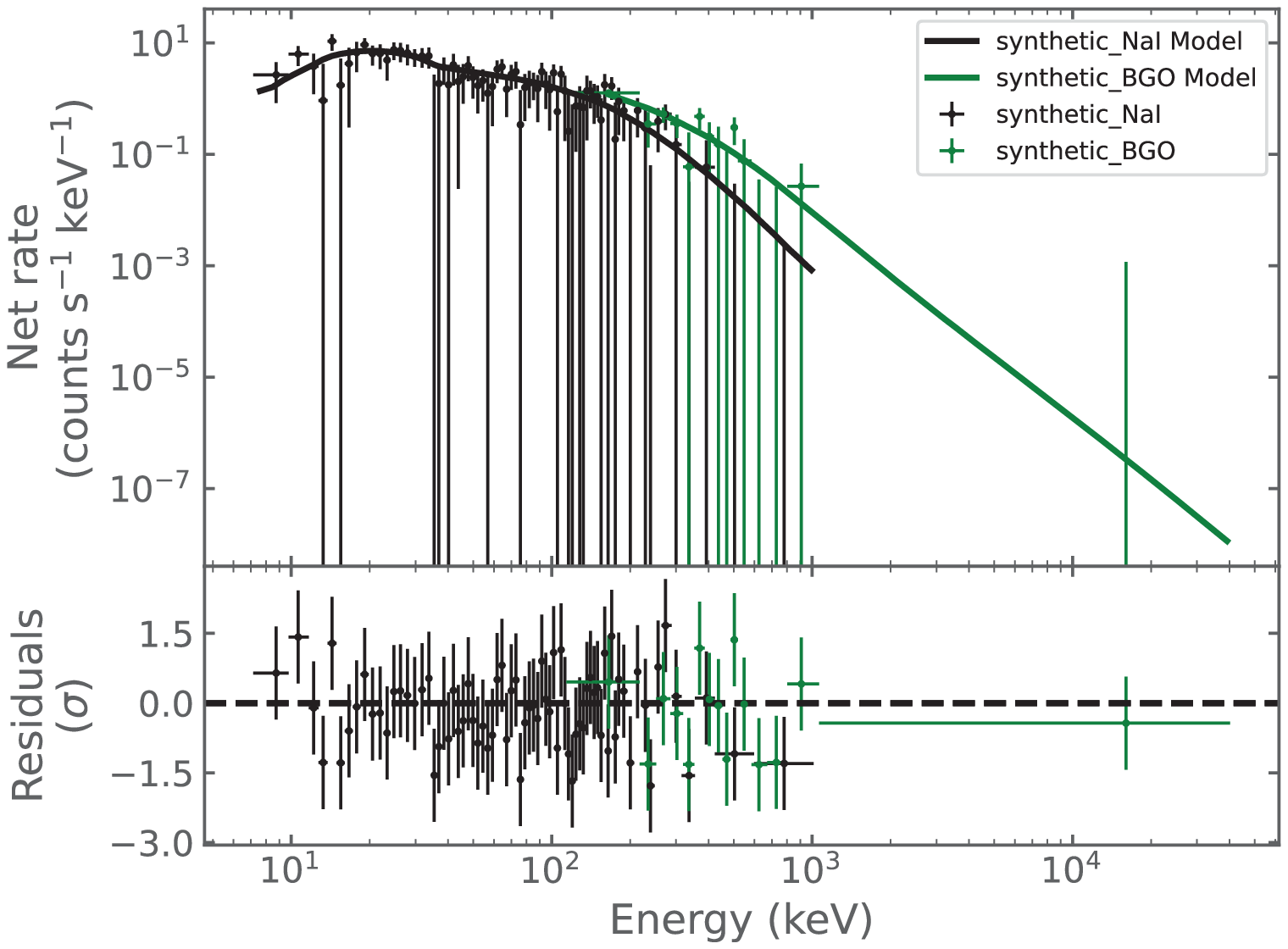}\\
\includegraphics[angle=0,scale=0.50]{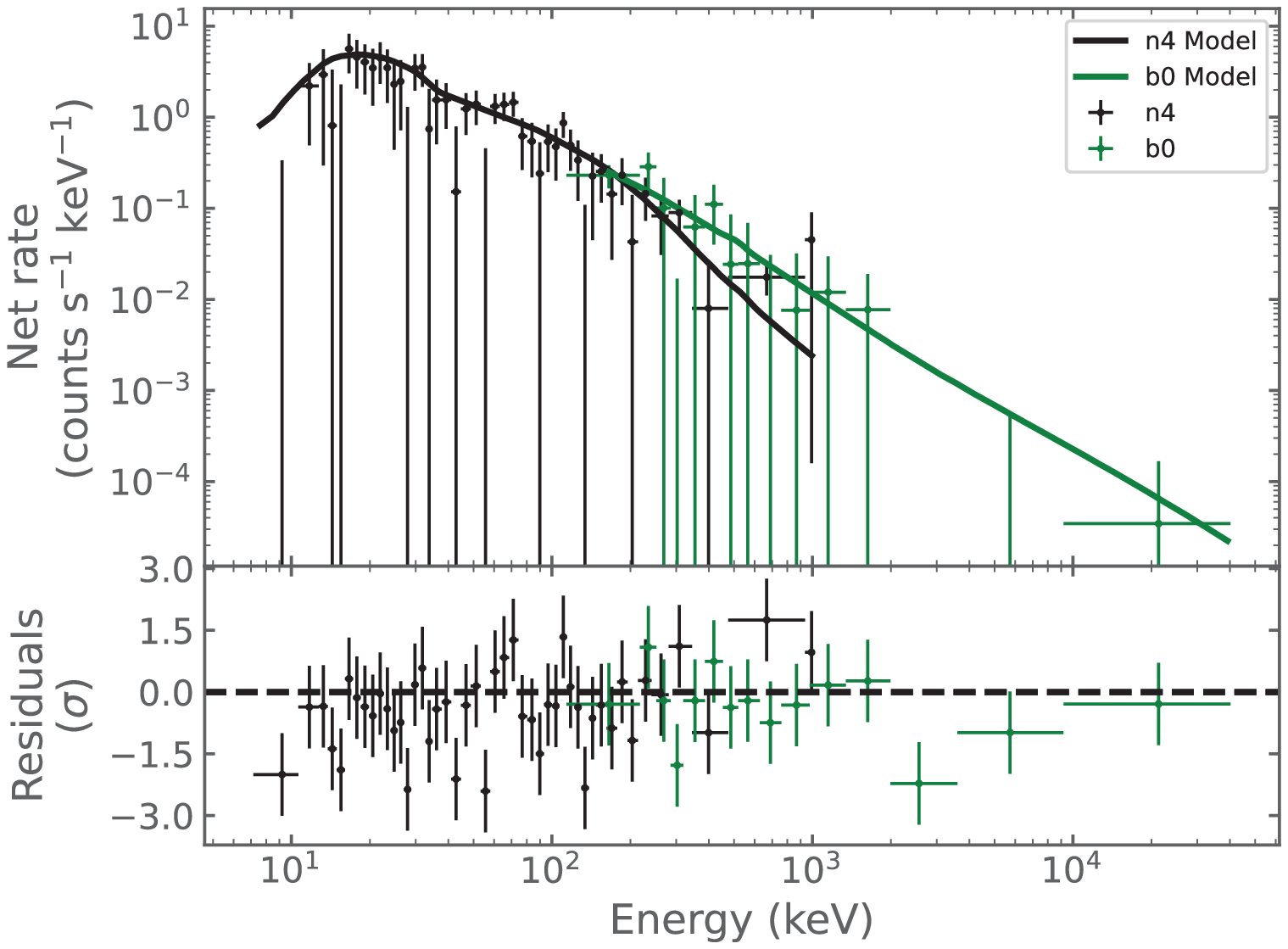}&
\includegraphics[angle=0,scale=0.50]{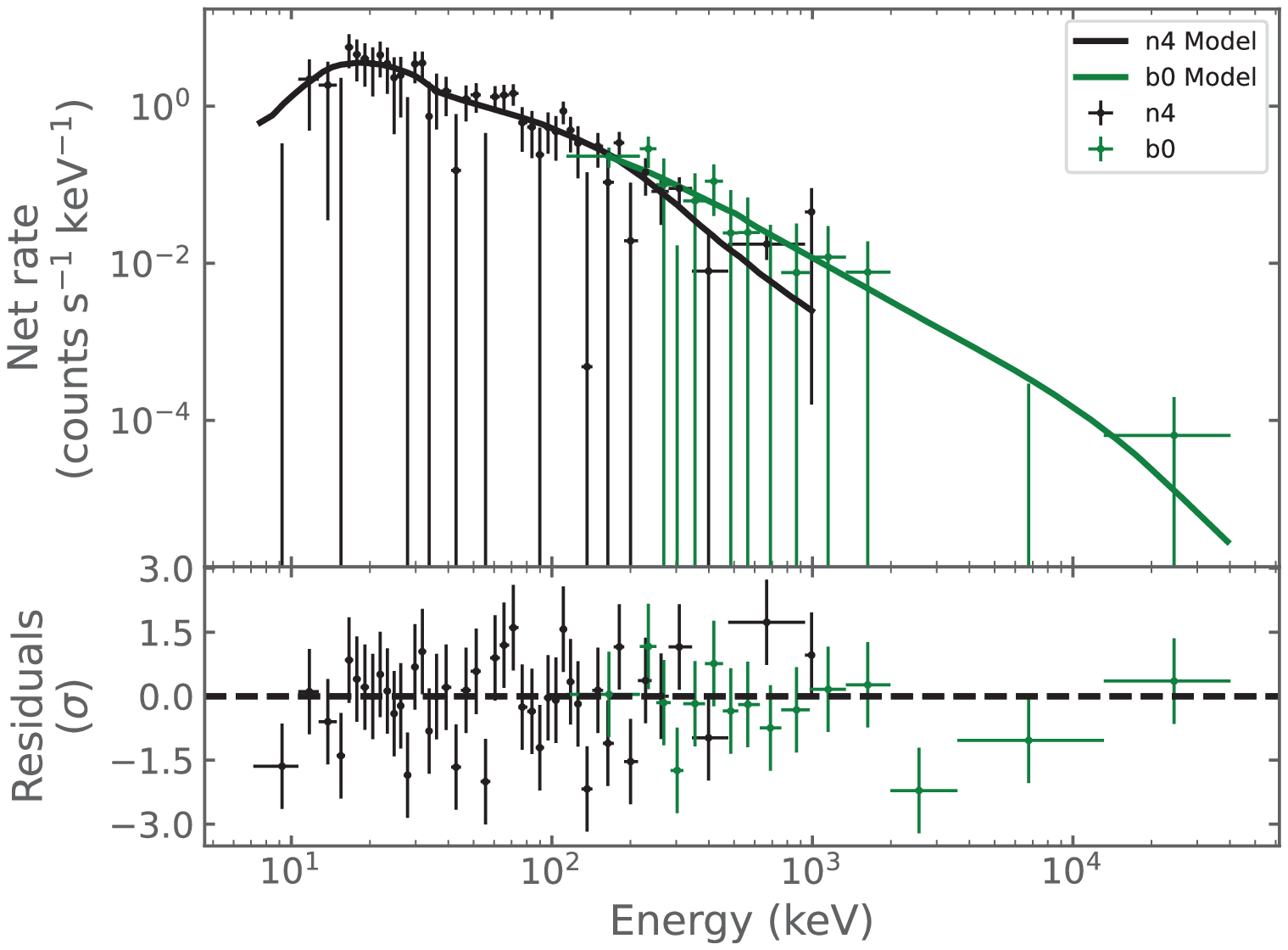}\\
\end{tabular}
\caption{Fitting results of the synthetic data (upper panels) and the observational data (bottom panels) in the 8~keV-40~MeV energy band,
where a PL spectral model and Band function are adopted in the left and right panels, respectively.}
\label{Appendix:MyFigE}
\end{figure}

%%0%%%%%-----------------------------------------------
\clearpage
\begin{figure}
\begin{tabular}{cc}
\includegraphics[angle=0,scale=0.3]{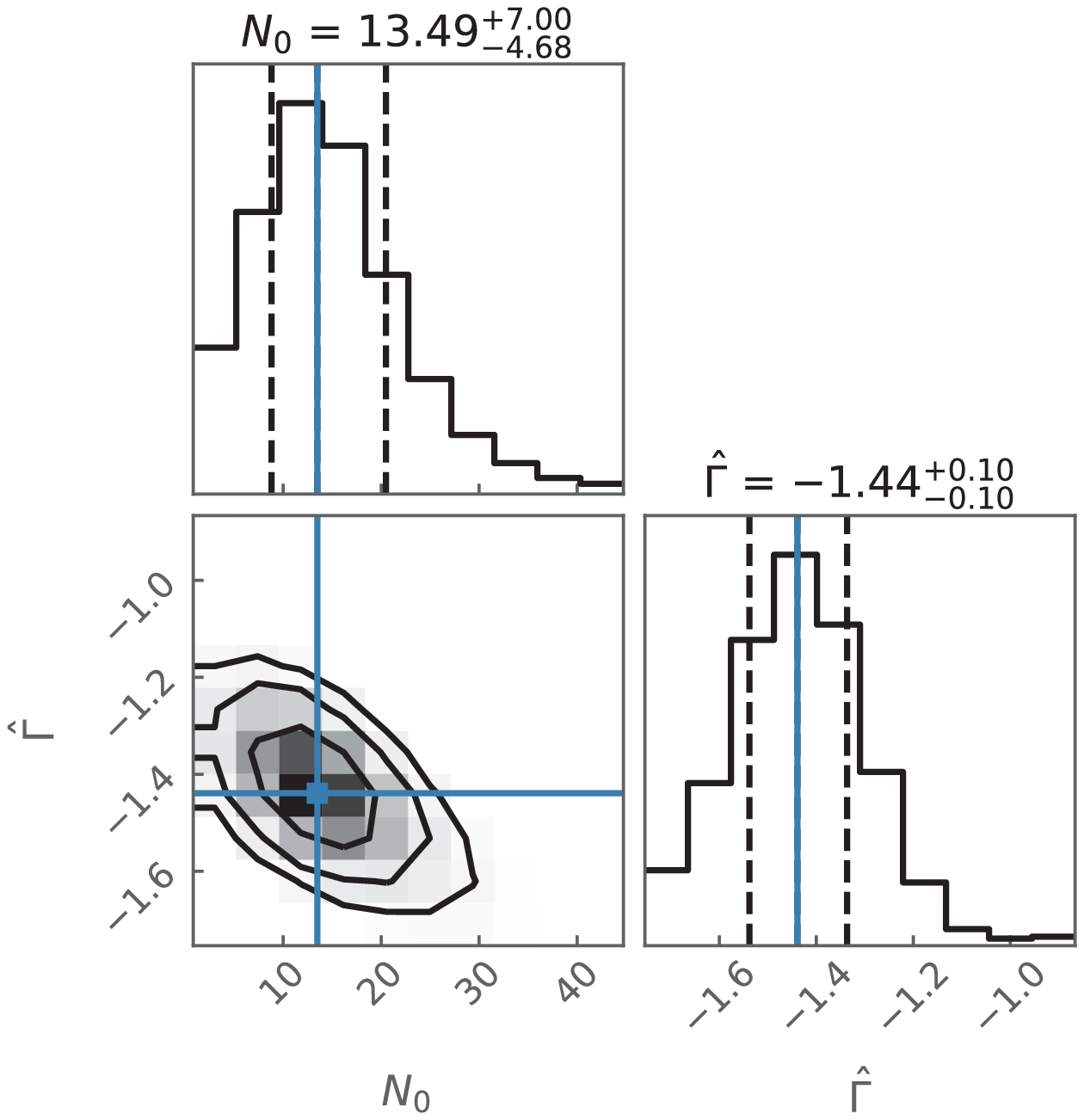}&
\includegraphics[angle=0,scale=0.3]{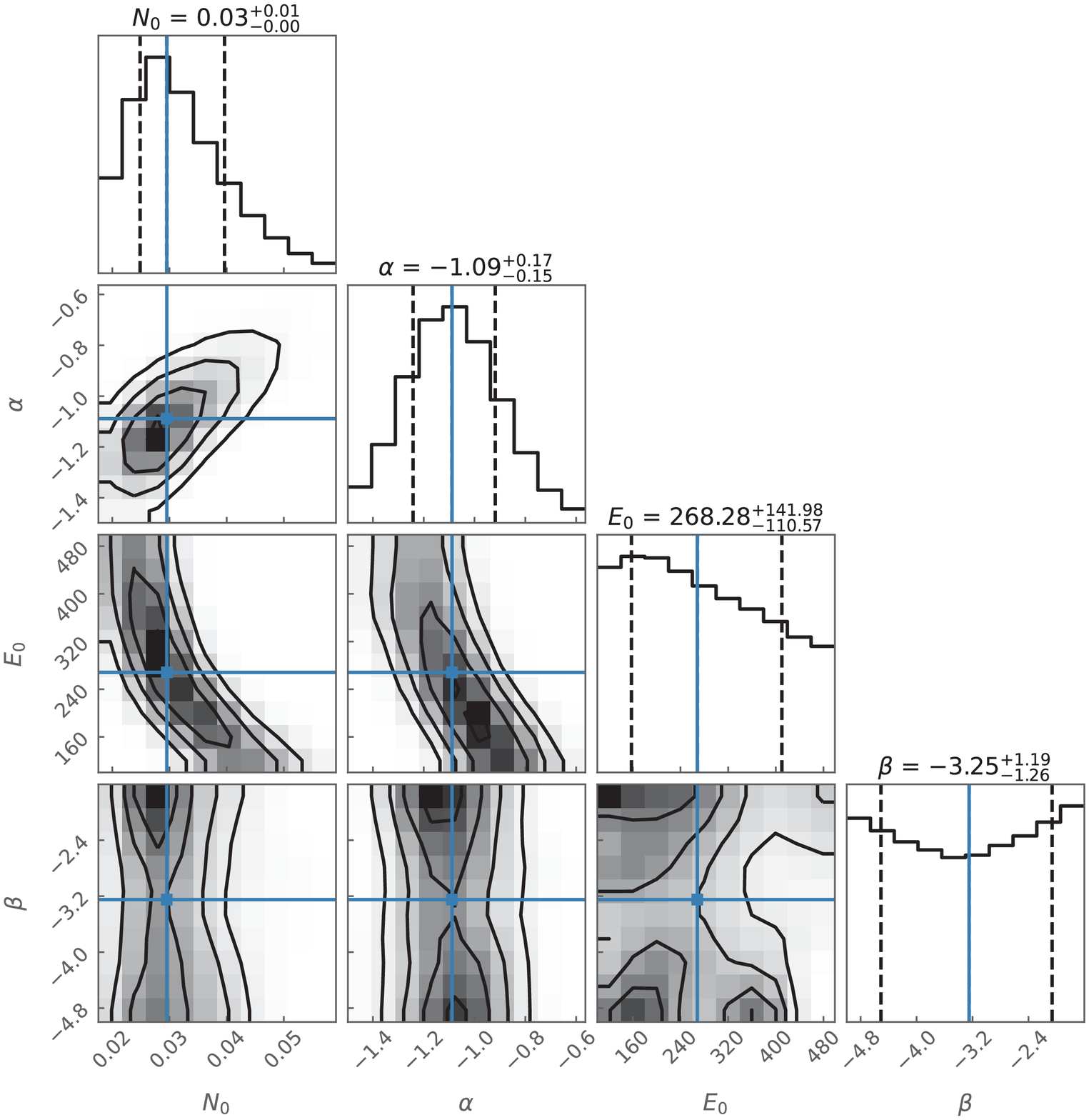}\\
\includegraphics[angle=0,scale=0.3]{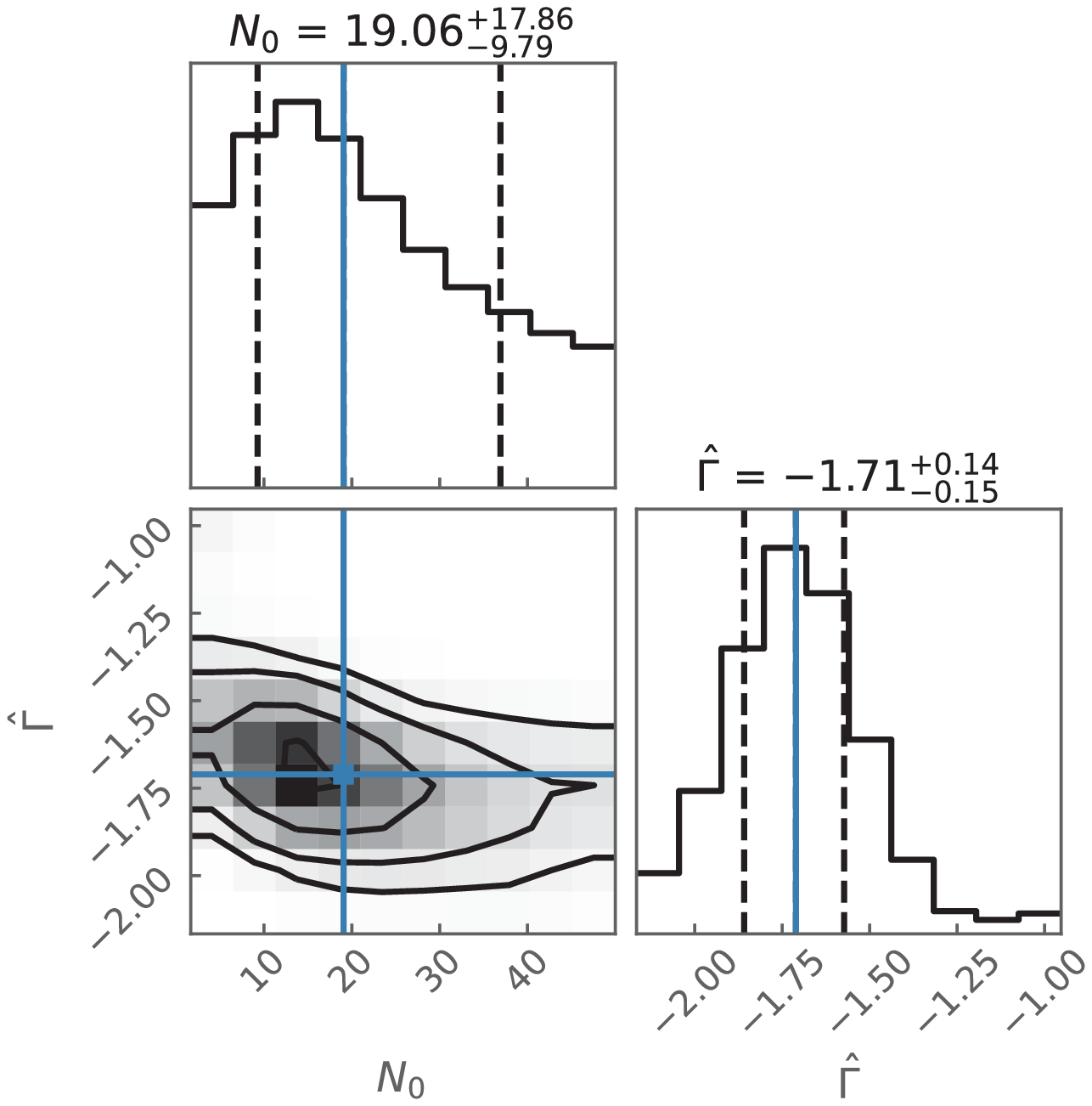} &
\includegraphics[angle=0,scale=0.3]{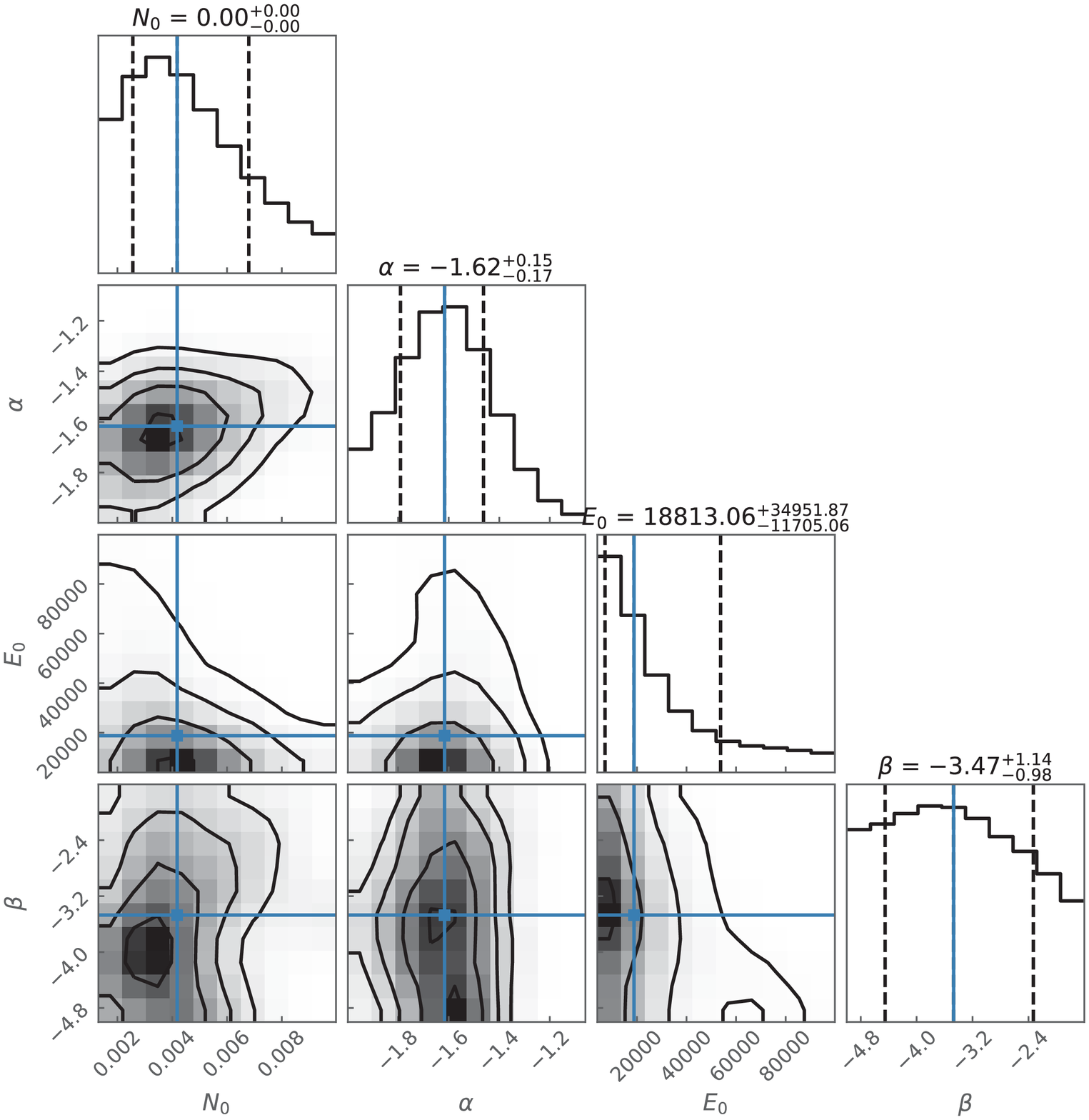}\\
\end{tabular}
\caption{Posterior probability density functions for
the physical parameters of the spectral fitting on the synthetic data (upper panels)
and the observational data (bottom panels) in the 15-150~keV energy band,
where a PL spectral model and Band function are adopted in the left and right panels, respectively.}
\label{Appendix:MyFigF}
\end{figure}
%%%%%%%%%%%%%%%%%%%%%%%%%%%%%%%%%%%%%%%%%%%%%%%%%%%%%%%%%%%%%%%%%%%%%%%%%%%%%%%%%%%%%%%%%%%%
\clearpage
\begin{table}
\tablewidth{500pt}
\caption{Spectral fitting results of simulation and observation of $[0, 5]$~s in GRB~200829A. }
\centering{
\begin{tabular}{ccccccc}
\hline\hline
Model  & $\alpha$ (or $\hat{\Gamma}$)& $\beta$ &$E_0  ({\rm keV})$ & $N_0$& Data sources  \\
\hline
PL  &  $-1.65_{-0.04}^{+0.04}$  &  - &  - &  $31.52_{-5.32}^{+6.14}$ &synthetic data (8~keV-40~MeV) \\
Band  &  $-1.16_{-0.11}^{+0.14}$ &  $-3.71_{-0.83}^{+0.88}$&  $276.67_{-71.06}^{+91.20}$&  $0.03_{-0.00}^{+0.01}$ & synthetic data (8~keV-40~MeV)  \\
PL  &  $-1.73_{-0.09}^{+0.08}$  &  - &  - &  $15.65_{-7.15}^{+11.32}$ &observational data (8~keV-40~MeV) \\
Band  &  $-1.61_{-0.11}^{+0.10}$ &  $-2.94_{-1.21}^{+0.85}$&  $11021.76_{-5533.27}^{+13229.91}$&  $0.00\pm0.00$ &observational data (8~keV-40~MeV)  \\
\hline
PL   & $-1.44_{-0.10}^{+0.10}$  &  - &  - &  $13.49_{-4.68}^{+7.00}$ &synthetic data (15-150~keV) \\
Band  &$-1.09_{-0.15}^{+0.17}$ &  $-3.25_{-1.26}^{+1.19}$&  $268.28_{-110.57}^{+141.98}$&  $0.03_{-0.00}^{+0.01}$ &synthetic data (15-150~keV)   \\
PL  & $-1.71_{-0.15}^{+0.14}$  &  - &  - &  $19.06_{-9.79}^{+17.86}$ &observational data (15-150~keV) \\
Band  &  $-1.62_{-0.17}^{+0.15}$ &  $-3.47_{-0.98}^{+1.14}$&  $18813.06_{-11705.06}^{+34951.87}$&  $0.00\pm0.00$ &observational data (15-150~keV)  \\
\hline
\end{tabular}}
\label{Appendix:MyTabD}
\end{table}

%%%%%%%%%%%%%%%%%%%%%%%%%%%%%%%%%%%%%%%%%%%%%%%%%%%%%%%%%%%%

\clearpage
%\bibliography{bibliography}

\end{document}